\documentstyle[12pt]{article}
\catcode`\@=11

\@addtoreset{equation}{section}
\def\theequation{\arabic{equation}}
\def\theequation{\thesection\arabic{equation}}

\newcommand{\be}{\begin{equation}}
\newcommand{\ee}{\end{equation}}
\newcommand{\ba}{\begin{eqnarray}}
\newcommand{\ea}{\end{eqnarray}}
\def\JGP#1#2#3{{\it J.~Geom.~Phys.} {\bf{B#1}} (19#2) #3}
\def\NPB#1#2#3{{\it Nucl.~Phys.} {\bf{B#1}} (19#2) #3}
\def\PLB#1#2#3{{\it Phys.~Lett.} {\bf{B#1}} (19#2) #3}
\def\PRD#1#2#3{{\it Phys.~Rev.} {\bf{D#1}} (19#2) #3}

\def\JHEP#1#2#3{{\it J. High Energy Phys.} {\bf#1} (19#2) #3}

\def\part{\partial}

\def\a{\alpha}

\def\d{\delta}

\def\p{\pi}

\def\t{\tau}

\def\@normalsize{\@setsize\normalsize{15pt}\xiipt\@xiipt
\abovedisplayskip 14pt plus3pt minus3pt%
\belowdisplayskip \abovedisplayskip
\abovedisplayshortskip  \z@ plus3pt%
\belowdisplayshortskip  7pt plus3.5pt minus0pt}
\def\small{\@setsize\small{13.6pt}\xipt\@xipt
\abovedisplayskip 13pt plus3pt minus3pt%
\belowdisplayskip \abovedisplayskip
\abovedisplayshortskip  \z@ plus3pt%
\belowdisplayshortskip  7pt plus3.5pt minus0pt
\def\@listi{\parsep 4.5pt plus 2pt minus 1pt
            \itemsep \parsep
            \topsep 9pt plus 3pt minus 3pt}}

\def\underline#1{\relax\ifmmode\@@underline#1\else
        $\@@underline{\hbox{#1}}$\relax\fi}
\@twosidetrue
\relax

\catcode`@=12

\catcode`\@=11

\def\section{\@startsection{section}{1}{\z@}{3.5ex plus 1ex minus
   .2ex}{2.3ex plus .2ex}{\large\bf}}
\def\thesection{\arabic{section}.}
\def\thesubsection{\arabic{section}.\arabic{subsection}}

\def\ps@headings{\def\@oddfoot{}\def\@evenfoot{}
\def\@oddhead{\hbox{}\hfill
        \makebox[.5\textwidth]{\raggedright\ignorespaces --\thepage{}--
        \hfill }}
\def\@evenhead{\@oddhead}
\def\subsectionmark##1{\markboth{##1}{}} }
\renewcommand{\subsection}[1]{\addtocounter{subsection}{1}
\vspace{2.5mm}\par\noindent {\em \thesubsection . #1}\par
 \vspace{0.5mm} }
\ps@headings

\catcode`\@=12

\relax

\def\figcap{\section*{Figure Captions\markboth
        {FIGURECAPTIONS}{FIGURECAPTIONS}}\list
        {Fig. \arabic{enumi}:\hfill}{\settowidth\labelwidth{Fig. 999:}
        \leftmargin\labelwidth
        \advance\leftmargin\labelsep\usecounter{enumi}}}
 \relax
\def\tablecap{\section*{Table Captions\markboth
        {TABLECAPTIONS}{TABLECAPTIONS}}\list
        {Table \arabic{enumi}:\hfill}{\settowidth\labelwidth{Table 999:}
        \leftmargin\labelwidth
        \advance\leftmargin\labelsep\usecounter{enumi}}}
 \relax
\def\reflist{\section*{References\markboth
        {REFLIST}{REFLIST}}\list
        {[\arabic{enumi}]\hfill}{\settowidth\labelwidth{[999]}
        \leftmargin\labelwidth
        \advance\leftmargin\labelsep\usecounter{enumi}}}
 \relax

\catcode`\@=11

\def\marginnote#1{}
\newcount\hour
\newcount\minute
\newtoks\amorpm
\hour=\time\divide\hour by60
\minute=\time{\multiply\hour by60 \global\advance\minute by-
\hour}
\edef\standardtime{{\ifnum\hour<12 \global\amorpm={am}%
    \else\global\amorpm={pm}\advance\hour by-12 \fi
    \ifnum\hour=0 \hour=12 \fi
    \number\hour:\ifnum\minute<100\fi\number\minute\the\amorpm}}
\edef\militarytime{\number\hour:\ifnum\minute<100\fi\number\minute}
\def\draftlabel#1{{\@bsphack\if@filesw {\let\thepage\relax
  \xdef\@gtempa{\write\@auxout{\string
    \newlabel{#1}{{\@currentlabel}{\thepage}}}}}\@gtempa
    \if@nobreak \ifvmode\nobreak\fi\fi\fi\@esphack}
     \gdef\@eqnlabel{#1}}
\def\@eqnlabel{}
\def\@vacuum{}
\def\draftmarginnote#1{\marginpar{\raggedright\scriptsize\tt#1}}
\def\draft{\oddsidemargin -.5truein
        \def\@oddfoot{\sl preliminary draft \hfil
        \rm\thepage\hfil\sl\today\quad\militarytime}
        \let\@evenfoot\@oddfoot \overfullrule 3pt
        \let\label=\draftlabel
        \let\marginnote=\draftmarginnote
   
\def\@eqnnum{(\theequation)\rlap{\kern\marginparsep\tt\@eqnlabel}%
\global\let\@eqnlabel\@vacuum}  }
\def\preprint{\twocolumn\sloppy\flushbottom\parindent 1em
        \leftmargini 2em\leftmarginv .5em\leftmarginvi .5em
        \oddsidemargin -.5in    \evensidemargin -.5in
        \columnsep 15mm \footheight 0pt
        \textwidth 250mmin      \topmargin  -.4in
        \headheight 12pt \topskip .4in
        \textheight 175mm
        \footskip 0pt
        
\def\@oddhead{\thepage\hfil\addtocounter{page}{1}\thepage}
        \let\@evenhead\@oddhead \def\@oddfoot{} \def\@evenfoot{}  }
\def\titlepage{\@restonecolfalse\if@twocolumn\@restonecoltrue\onecolumn
     \else \newpage \fi \thispagestyle{empty}\c@page\z@
        \def\thefootnote{\fnsymbol{footnote}} }
\def\endtitlepage{\if@restonecol\twocolumn \else  \fi
        \def\thefootnote{\arabic{footnote}}
        \setcounter{footnote}{0}}  
\catcode`@=12
\relax

\def\ps@headings{\def\@oddfoot{}\def\@evenfoot{}
\def\@oddhead{\hbox{}\hfill
        \makebox[.5\textwidth]{\raggedright\ignorespaces --\thepage{}--
        \hfill }}
\def\@evenhead{\@oddhead}
\def\subsectionmark##1{\markboth{##1}{}} }

\ps@headings

\relax

\def\firstpage#1#2#3#4#5#6{
\begin{document}


\begin{titlepage}
\nopagebreak
\title{\begin{flushright}
        \vspace*{-1.8in}
        {\normalsize CPTH-S725.0799}\\[-10mm]
        {\normalsize DFF/340/07/99}\\[-10mm]
       {\normalsize LPT-ORSAY 99/58}\\[-10mm]
        {\normalsize ROM2F-99/22}\\[-10mm]
        {\normalsize hep-th/9907184}\\[-4mm]
\end{flushright}
\vfill {#3}}
\author{\large #4 \\[1.0cm] #5}
\maketitle
\vskip -9mm     
\nopagebreak 
\begin{abstract} {\noindent #6}
\end{abstract}
\vfill
\begin{flushleft}
\rule{16.1cm}{0.2mm}\\[-4mm]
$^{\star}${\small Research supported in part by the EEC under TMR contract 
ERBFMRX-CT96-0090.}\\[-4mm] 
$^{\dagger}${CNRS-UMR-7644.}\\[-4mm]
$^{\ddagger}${\small Laboratoire associ{\'e} au CNRS-URA-D0063.}\\
\today
\end{flushleft}
\thispagestyle{empty}
\end{titlepage}}

\date{}
\firstpage{3118}{IC/95/34} {\large\bf Open Descendants of $Z_2 \times
Z_2$ Freely-Acting Orbifolds}  
{I. Antoniadis$^{\,a}$, G. D'Appollonio$^{\,b}$, E. Dudas$^{\,c}$ and 
A. Sagnotti$^{\,a,d}$} 
{\small\sl
$^a$ Centre de Physique Th{\'e}orique$^\dagger$,  Ecole Polytechnique, 
{}F-91128 Palaiseau\\[-3mm]
\small\sl$^{b}$ Dipartimento di Fisica, Universit\`a di Firenze\\[-4mm]
\small\sl INFN-Sezione di Firenze\\[-4mm]
\small\sl Largo Enrico Fermi 2, 50125 Firenze \ ITALY
\\[-3mm]
\small\sl $^c$  LPT$^\ddagger$, B{\^a}t. 210, Univ. Paris-Sud, F-91405 Orsay\\[-3mm] 
\small\sl$^{d}$ Dipartimento di Fisica, Universit\`a di Roma ``Tor Vergata''\\[-4mm]
\small\sl INFN,
 Sezione di Roma ``Tor Vergata''\\[-4mm]
\small\sl Via della Ricerca Scientifica 1, 00133 Roma \ ITALY} 
{We discuss $Z_2 \times Z_2$ orientifolds where
the orbifold twists are accompanied by shifts on momentum or 
winding lattice states. The models contain variable numbers of D5
branes, whose massless (and, at times, even massive) modes 
have variable numbers of
supersymmetries. We display new type-I models with partial supersymmetry
breaking $N=2 \rightarrow N=1$, $N=4 \rightarrow N=1$ and
$N=4 \rightarrow N=2$.
 The geometry of these models is rather rich: the shift operations 
create brane multiplets related by orbifold transformations that
support gauge groups of reduced rank. Some of the 
models are deformations of six-dimensional supersymmetric type-I
models, while others have dual M-theory descriptions. }
\section{Introduction}

Freely acting orbifolds are a useful tool for connecting string vacua
with different numbers of supersymmetries. The resulting models
provide string realizations of the Scherk-Schwarz mechanism 
\cite{ss,stringss,kk}, where
 additional supersymmetries may be recovered in an appropriate 
decompactification limit. If the original model has extended
supersymmetry, the corresponding 
BPS states, still present in the vacuum 
with lower supersymmetry, have
mass shifts determined by the spontaneous breaking. 
As a result, duality relations generally
continue to hold by the adiabatic argument, and the dynamics of the
new vacua simplifies greatly \cite{vw}.

Type I models may be directly linked \cite{carg} to models of oriented
closed strings, and in particular to those considered in \cite{kk}. 
They allow an interesting new possibility compared to the heterotic and
type II models studied previously \cite{ads,adds}: branes orthogonal to
the coordinate used for the breaking have full supersymmetry for their massless
excitations at tree level, and feel the breaking only through
radiative corrections. This
phenomenon, generically non-perturbative on the heterotic side, 
is of some interest in the current literature on brane
Kaluza-Klein scenarios, where it is
usually referred to as ``brane supersymmetry'' \cite{kt}. From a more
technical perspective, the closed sector of these models involves asymmetric
projections, so that the construction of the open descendants presents
some peculiar features. Asymmetric models have recently been
studied in a different context, since they provide an interesting route
toward a small or vanishing cosmological constant
\cite{ks}. The corresponding open descendants, studied in \cite{bg},
display an even more peculiar feature: all massive brane modes, not
only the massless ones, can have an extended supersymmetry \cite{kt}.

In this work we provide new examples of this class of models in the
context of $Z_2 \times Z_2$ orbifolds  where the conventional
 projections are accompanied by 
$Z_2$-valued momentum or winding lattice shifts. 
Their open descendants exhibit $N=4 \rightarrow N=1$ or 
$N=2 \rightarrow N=1$ partial supersymmetry breaking, and involve
interesting D-brane configurations, with a 
rich geometrical structure, that we describe in some detail.
These models display enhanced supersymmetry for the massless modes
on the branes, and in some
cases also for their massive excitations, as the asymmetric
orbifolds discussed in \cite{bg}. The presence of extended
supersymmetry in some of the branes, however, has a drawback in the
present construction, since the models we exhibit here are not chiral. 

An important new feature is that the branes typically arrange themselves in 
multiplets of images that are interchanged by some of the
orbifold operations. The components of a multiplet share the corresponding 
Chan-Paton charges and, as a result,
the rank of the corresponding gauge groups is reduced, 
as in models with a quantized NS antisymmetric tensor \cite{Bab}.
Here the displacement of branes that results in the creation of the
multiplets is generally unavoidable, so that even the most symmetric
configuration typically involves several images. Moreover, the
image branes play a crucial role in the local tadpole cancellation along
the (shifted) directions transverse to the branes. More precisely, as
we will discuss in detail in explicit models, a doublet of image branes 
ensures the local cancellation of tadpoles introduced 
in one transverse direction
by its two orientifold plane charges, while a quadruplet of
image branes ensures local tadpole cancellation in two transverse (shifted)
directions. These conditions are effective if the various types of
branes do not intersect, a property shared by all our configurations.
As a result, in most of our models the local tadpole conditions
are automatically satisfied, thanks to the images that appear
since the branes are generally forced 
away from their fixed points. Other brane configurations with less
supersymmetry are also possible, but they do not satisfy local tadpole
cancellation. Compared with toroidal
examples, where local tadpoles require that
the gauge group, of rank 16, be broken by Wilson lines \cite{pw} , 
in our examples a set of $m$ image branes share the same gauge 
group that, however, has a rank reduced to $16/m$. 

This paper is organized as follows.
In Section 2 we describe the structure of the supersymmetric 
$Z_2 \times Z_2$ model without discrete torsion\footnote{Discrete 
torsion introduces peculiar new features in the open descendants, and 
will be discussed elsewhere \cite{dt}.}. In Section 3 we define the ``shift 
orientifolds'' and describe
qualitatively their main properties, the structure of the
corresponding D-brane configurations and identify various
instances of 
brane supersymmetry. In Section 4 we present the conformal field
theory description and the open string partition functions for branes away
from fixed points in simple examples, while in Section 5 we describe a model
with $N=2 \rightarrow N=1$ breaking that displays rather clearly
all the salient features
of our constructions. Section 6, 7 and 8 contain a description
of models with two, one and zero sets of D5 branes, including their 
limiting behaviors for large or small radii where extended
supersymmetry is recovered, and their relation
to M-theory compactifications. 
Finally, Section 9 contains the summary of our results and our 
conclusions.


\section{Supersymmetric $Z_2 \times Z_2$ models}

There are two classes of supersymmetric $Z_2 \times Z_2$ models,
that differ from 
one another because of the presence of ``discrete torsion''
\cite{disc}. In the
partition function, this
is a sign associated to an independent modular orbit,
with crucial effects both on the massless field content
and on the structure of the descendants. These orbifolds are singular
limits of Calabi-Yau spaces,
with Hodge numbers $(3,51)$ and $(51,3)$ respectively. 

The torus amplitudes for the supersymmetric $Z_2 \times Z_2$ models are
\ba
{\cal T}\!\!\!&=&\!\!{1 \over 4} \biggl\{ |T_{oo}|^2 \Lambda_1 \Lambda_2 \Lambda_3
+|T_{og}|^2  \Lambda_1 
|{4\eta^2 \over \theta_2^2}|^2 + |T_{oh}|^2 \Lambda_3 
|{4\eta^2 \over \theta_2^2}|^2+
|T_{of}|^2 \Lambda_2 
|{4\eta^2 \over \theta_2^2}|^2 \nonumber \\
&+& |T_{go}|^2 \Lambda_1
|{4\eta^2 \over \theta_4^2}|^2+ |T_{gg}|^2  \Lambda_1 
|{4\eta^2 \over \theta_3^2}|^2 
+ |T_{ho}|^2 \Lambda_3
|{4\eta^2 \over \theta_4^2}|^2 +|T_{hh}|^2 
\Lambda_3 |{4\eta^2 \over \theta_3^2}|^2  \nonumber \\
&+& |T_{fo}|^2 
\Lambda_2 |{4 \eta^2 \over \theta_4^2}|^2 + 
|T_{ff}|^2 \Lambda_2 |{4 \eta^2 \over \theta_3^2}|^2  \nonumber \\
&+& \epsilon \left( |T_{gh}|^2 + |T_{gf}|^2 + |T_{fg}|^2 + |T_{fh}|^2 + |T_{hg}|^2
+ |T_{hf}|^2 \right) |{8\eta^3 \over {\theta_2 \theta_3 \theta_4}}|^2 
\biggr\} \quad  , 
\label{s1} 
\ea
where $\Lambda_1$, $\Lambda_2$ and $\Lambda_3$ denote the three
lattice sums associated to the three internal tori $T_{45}$, $T_{67}$
and $T_{89}$
of the compactification and $\epsilon = \pm 1$. Here, as in all
following amplitudes, we are leaving the contributions of transverse
bosons implicit. The choice
$\epsilon=1$ defines the model without discrete torsion, while
the choice $\epsilon=-1$ defines the model with discrete torsion.
The model
with discrete torsion is quite interesting, since its descendants 
contain chiral fermions, but presents some difficulties, since some of
the ``untwisted'' tadpoles can not be eliminated in the usual way, 
and will be discussed elsewhere.

As usual, the arguments depend on $q = \exp( 2 \pi i \tau)$ and its
conjugate, where $\tau$ is the modulus of the torus.
For later convenience, we have expressed the torus amplitude
in terms of the 16 quantities $T_{ij}$ $(i=o,g,h,f)$:
\ba
T_{io} &=&  \tau_{io} +  \tau_{ig} + \tau_{ih} + \tau_{if} \quad , \qquad
T_{ig} =  \tau_{io} +  \tau_{ig} - \tau_{ih} - \tau_{if} \quad , \nonumber \\
T_{ih} &=&  \tau_{io} -  \tau_{ig} + \tau_{ih} - \tau_{if} \quad , \qquad
T_{if} =  \tau_{io} -  \tau_{ig} - \tau_{ih} + \tau_{if} \quad ,
\label{s2}
\ea
where the $16$ $Z_2 \times Z_2$ characters $\tau_{ij}$ are \cite{erice}
\ba
\tau_{oo}&=&V_2O_2O_2O_2+O_2V_2V_2V_2-S_2S_2S_2S_2-C_2C_2C_2C_2 \ , \nonumber \\
\tau_{og}&=&O_2V_2O_2O_2+V_2O_2V_2V_2-C_2C_2S_2S_2-S_2S_2C_2C_2 \ , \nonumber \\
\tau_{oh}&=&O_2O_2O_2V_2+V_2V_2V_2O_2-C_2S_2S_2C_2-S_2C_2C_2S_2 \ , \nonumber \\
\tau_{of}&=&O_2O_2V_2O_2+V_2V_2O_2V_2-C_2S_2C_2S_2-S_2C_2S_2C_2 \ , \nonumber \\
\tau_{go}&=&V_2O_2S_2C_2+O_2V_2C_2S_2-S_2S_2V_2O_2-C_2C_2O_2V_2 \ , \nonumber \\
\tau_{gg}&=&O_2V_2S_2C_2+V_2O_2C_2S_2-S_2S_2O_2V_2-C_2C_2V_2O_2 \ , \nonumber \\
\tau_{gh}&=&O_2O_2S_2S_2+V_2V_2C_2C_2-C_2S_2V_2V_2-S_2C_2O_2O_2 \ , \nonumber \\
\tau_{gf}&=&O_2O_2C_2C_2+V_2V_2S_2S_2-S_2C_2V_2V_2-C_2S_2O_2O_2 \ , \nonumber \\
\tau_{ho}&=&V_2S_2C_2O_2+O_2C_2S_2V_2-C_2O_2V_2C_2-S_2V_2O_2S_2 \ , \nonumber \\
\tau_{hg}&=&O_2C_2C_2O_2+V_2S_2S_2V_2-C_2O_2O_2S_2-S_2V_2V_2C_2 \ , \nonumber \\
\tau_{hh}&=&O_2S_2C_2V_2+V_2C_2S_2O_2-S_2O_2V_2S_2-C_2V_2O_2C_2 \ , \nonumber \\
\tau_{hf}&=&O_2S_2S_2O_2+V_2C_2C_2V_2-C_2V_2V_2S_2-S_2O_2O_2C_2 \ , \nonumber \\
\tau_{fo}&=&V_2S_2O_2C_2+O_2C_2V_2S_2-S_2V_2S_2O_2-C_2O_2C_2V_2 \ , \nonumber \\
\tau_{fg}&=&O_2C_2O_2C_2+V_2S_2V_2S_2-C_2O_2S_2O_2-S_2V_2C_2V_2 \ , \nonumber \\
\tau_{fh}&=&O_2S_2O_2S_2+V_2C_2V_2C_2-C_2V_2S_2V_2-S_2O_2C_2O_2 \ , \nonumber \\
\tau_{ff}&=&O_2S_2V_2C_2+V_2C_2O_2S_2-C_2V_2C_2O_2-S_2O_2S_2V_2 \ , 
\label{s3}
\ea
with  $O_2,V_2,S_2,C_2$ the four $O(2)$ level-one characters. The
ordering  of the four factors refers to the eight transverse
dimensions of space time and, in particular, the
first factor is
associated to the two transverse space-time directions. The notation
$(o,g,h,f)$ reflects the four operations in $Z_2 \times Z_2$. 
Aside from the identity $o$, these act as $\pi$-rotations on two of the three
internal tori, namely
\be
g: (+,-,-) \quad , \qquad f: (-,+,-) \quad , \qquad h: (-,-,+) \quad .
\label{s35}
\ee

One can now follow the usual procedure and build the three additional
amplitudes that, together with (\ref{s1}), determine the
spectrum  of the open descendants:
Klein-bottle ${\cal K}$, annulus ${\cal A}$ and M\"obius
strip ${\cal M}$. There are actually a number of distinct choices for
${\cal K}$, that correspond to different choices for three signs
$(\epsilon_1$,$\epsilon_2$,$\epsilon_3$) that define
the world-sheet projection for twisted states. In all cases these
signs are related to the parameter $\epsilon$ introduced previously,
since they must satisfy
\be
\epsilon_1 \ \epsilon_2 \ \epsilon_3 \ = \epsilon \quad .
\label{s4}
\ee

Thus, the models with discrete torsion require an {\it odd} number of
negative signs, while the models without discrete torsion require an
{\it even} number of negative signs.
Let us now introduce the restrictions of the
full lattice sums $\Lambda_i$ to their momentum and winding 
sublattices. These will
be denoted by $P_i$ and $W_i$, where the index identifies the
corresponding two-torus. In later Sections, as in \cite{ads,adds}, we
shall also need sums with alternating signs in one direction of the 
two-torii, that we shall denote
succinctly as $(-1)^m P$ or $(-1)^n W$. 
In terms of these quantities the Klein-bottle
amplitude in the direct channel is
\ba
{\cal K} &=&\frac{1}{8} \biggl\{ ( P_1 P_2 P_3 + P_1 W_2 W_3 + W_1 P_2
W_3 + W_1 W_2 P_3 ) T_{oo} \nonumber \\ & &+ 
2 \times 16 \bigl[\epsilon_1 (P_1 + \epsilon W_1 ) T_{go} 
+  \epsilon_2 (P_2 + \epsilon W_2 ) T_{fo}
+ \epsilon_3 (P_3 + \epsilon W_3 ) T_{ho} \bigr] \left( 
\frac{\eta}{\theta_4} \right)^2 \biggr\} \quad .
\label{s5}
\ea
The discrete torsion 
has a crucial effect on this term. Indeed, while for $\epsilon=1$ the
massless twisted contributions are diagonal combinations of the
$\tau$'s, and describe $N=2$ vector multiplets, for $\epsilon=-1$ they are
off-diagonal ones that describe (chiral-linear) hypermultipets, and therefore
do not contribute to the Klein bottle amplitude. This feature reflects
itself in the nature of the terms
$(P_i + \epsilon W_i)$ in  ${\cal K}$. While for $\epsilon=-1$ these
have no massless contributions, for $\epsilon=+1$ they do, and one has
additional options, related to the $\epsilon_i$. For instance, starting from
the original twisted $N=2$ vector multiplets, in the
Klein-bottle projection $\epsilon_1=1$  selects $N=1$
chiral multiplets, while $\epsilon_1=-1$ selects vector
multiplets.

An $S$ transformation turns this expression into the corresponding
vacuum-channel amplitude
\ba
\tilde{\cal K}\!\!&=&\!\! \frac{2^5}{8} \biggl\{ \bigl( v_1v_2v_3 W^e_1 W^e_2 W^e_3
+  \frac{v_1}{v_2 v_3} W_1^e P_2^e P_3^e + \frac{v_2}{v_1 v_3} P_1^e W_2^e
P_3^e + \frac{v_3}{v_1 v_2} P_1^e P_2^e W_3^e \bigr) T_{oo} \nonumber \\
\!\!\!\!\! &+&\!\!\!\! 2 \bigl[\epsilon_1 (v_1 W_1^e + \epsilon \frac{P_1^e}{v_1}) T_{og} 
+  \epsilon_2  (v_2 W_2^e + \epsilon \frac{P_2^e}{v_2}) T_{of}
+ \epsilon_3  (v_3 W_3^e + \epsilon \frac{P_3^e}{v_3}) T_{oh} \bigr] 
\left( \frac{2
\eta}{\theta_2} \right)^2 \biggr\} \, ,
\label{s6}
\ea
where the superscript $e$ stands for the usual restriction of the
lattice terms to their even subsets.
The structure of this term relates the three signs $\epsilon_i$
to the discrete torsion $\epsilon=\pm 1$.  This is neatly displayed by confining
the attention to the terms at the origin of the lattices, that are
to group into perfect squares. Indeed, the corresponding amplitude,
\ba
\tilde{\cal K}_0 &=& \frac{2^5}{8} \biggl\{ \ \biggl( \sqrt{v_1v_2v_3} +
 \epsilon_1 \sqrt{\frac{v_1}{v_2 v_3}} + 
\epsilon_2 \sqrt{\frac{v_2}{v_1 v_3}}  + \epsilon_3
\sqrt{\frac{v_3}{v_1 v_2}} \biggr)^2 \tau_{oo} \nonumber \\ 
& &+  \biggl( \sqrt{v_1v_2v_3} +
 \epsilon_1 \sqrt{\frac{v_1}{v_2 v_3}} - 
\epsilon_2 \sqrt{\frac{v_2}{v_1 v_3}}  - \epsilon_3
\sqrt{\frac{v_3}{v_1 v_2}} \biggr)^2 \tau_{og} \nonumber \\
& &+ \biggl( \sqrt{v_1v_2v_3} -
 \epsilon_1 \sqrt{\frac{v_1}{v_2 v_3}} + 
\epsilon_2 \sqrt{\frac{v_2}{v_1 v_3}}  - \epsilon_3
\sqrt{\frac{v_3}{v_1 v_2}} \biggr)^2 \tau_{of} \nonumber \\
&&+ \biggl( \sqrt{v_1v_2v_3} -
 \epsilon_1 \sqrt{\frac{v_1}{v_2 v_3}} - 
\epsilon_2 \sqrt{\frac{v_2}{v_1 v_3}}  + \epsilon_3
\sqrt{\frac{v_3}{v_1 v_2}} \biggr)^2 \tau_{oh} 
 \ \biggr\}
\label{s7}
\ea
displays four independent squared reflection coefficients, related to
the 9-brane and to the three types of 5-branes, {\it only} if
eq. (\ref{s4}) is satisfied. It should be appreciated that the constraint
(\ref{s4}) implies that, in the presence of discrete
torsion, at least one of the orientifold charges is reversed.
This is the problem we mentioned at the beginning.
Here we confine our attention to the choice 
$\epsilon_1=\epsilon_2=\epsilon_3=1$ for the model without discrete
torsion.

It is instructive to review the structure of the open descendants for
the model without discrete torsion. The orbifold breakings are not
allowed in this case: the twisted terms involve diagonal combinations
of the (non-self-conjugate) characters $\tau_{ij}$, that cannot flow 
in the tube channel.
The allowed terms
describe the familiar $NN$, $ND$ and $DD$ strings or, in brane
language, the $99$, $5_i9$ and $5_i5_i$ terms, together with additional,
more peculiar, $5_i5_j$ strings. For the sake of simplicity,
let us also refrain from introducing Wilson lines,
insert all $5$-branes at the origin and exclude a quantized 
$NS$ $B_{ab}$, in order
to obtain a relatively simple model with a maximal gauge group.  

The annulus amplitude for the model without discrete torsion is
\ba
{\cal A} &=& \frac{1}{8} \biggl\{ \bigl( N^2 \ P_1 P_2 P_3 + 
D_1^2 \ P_1 W_2 W_3 
 + D_2^2 \ W_1 P_2 W_3 + D_3^2 \ W_1 W_2 P_3 \bigr) T_{oo} \nonumber \\
&+& 2 \bigl( N D_1 \  P_1 T_{go} + N D_2 \ P_2 T_{fo} + N D_3 \ P_3 T_{ho}
\bigr) \left( \frac{
\eta}{\theta_4} \right)^2 \nonumber \\
&+& 2 \bigl(  D_2 D_3 \ W_1 T_{go} + 
D_1 D_3 \ W_2 T_{fo} + D_1 D_2 \ W_3 T_{ho} \bigr) \left( \frac{
\eta}{\theta_4} \right)^2  \biggl\}
\quad .
\label{s8}
\ea
Aside from the standard $NN$ terms associated with the 9-branes, there are 
indeed three
types of standard $D_i^2$ terms, and corresponding $ND_i$ terms. 
Each of these $D_iD_i$ strings, however, is only 
confined within one of the three two-tori, but is free to move at will on
it. In addition, there are three other types of strings of
a more peculiar nature, where Dirichlet and Neumann conditions
are entangled. The need for all these types of strings can easily be
appreciated 
by noticing that, relatively to the three internal two-tori,  the 
three twists of the $Z_2 \times Z_2$ orbifold result in open string ends with
boundary conditions of four types: $NNN$, $NDD$, $DND$ and
$DDN$. Combining these boundaries in all possible ways yields all
types of open strings present in the model. Aside 
from the familiar $NN$, $ND_i$ and $D_iD_i$ strings, the spectrum 
thus includes 
three additional types of open strings, that are $DD$ with respect
to a torus and 
$ND$ with respect to the others. For instance, the combination
of $NDD$ and $DND$ is a string of this type, DD in the
third torus and $ND$ in the first two.

An $S$ transformation turns (\ref{s8}) into the corresponding
vacuum-channel amplitude
\ba
\tilde{\cal A} &=& \frac{2^{-5}}{8} \biggl\{
\bigl( N^2 \ v_1 v_2 v_3 W_1 W_2 W_3 + \frac{D_1^2 v_1}{v_2 v_3} \ W_1 P_2 P_3 
 +  \frac{D_2^2 v_2}{v_1 v_3} P_1 W_2 P_3 +  \frac{D_3^2 v_3}{v_1
v_2} P_1 P_2 W_3 \bigr) T_{oo} \nonumber \\
& &+ 2 \bigl( N D_1 \ v_1 W_1 T_{og} + N D_2 \ v_2 W_2 T_{of} + 
N D_3 \ v_3 W_3 T_{oh}
\bigr) \left( \frac{2
\eta}{\theta_2} \right)^2 \nonumber \\
& &+ 2 \bigl( D_2 D_3 \
\frac{1}{v_1} P_1 T_{og} + 
D_1 D_3 \ \frac{1}{v_2} P_2 T_{of} + D_1 D_2 \ \frac{1}{v_3} P_3 T_{oh} \bigr) \left( \frac{2
\eta}{\theta_2} \right)^2 \biggr\}  \quad ,
\label{s9}
\ea
whose neat structure can be exhibited
confining the attention to the terms at the origin of the lattice sums,
that as usual rearrange themselves into perfect squares. Indeed,
\ba
\tilde{\cal A}_0 &=&
\frac{2^{-5}}{8} \biggl\{ \ \bigl( N \sqrt{v_1v_2v_3} +
 D_1 \sqrt{\frac{v_1}{v_2 v_3}} + 
D_2 \sqrt{\frac{v_2}{v_1 v_3}}  + D_3
\sqrt{\frac{v_3}{v_1 v_2}} \bigr)^2 \tau_{oo} \nonumber \\ 
& &+  \bigl( N \sqrt{v_1v_2v_3} +
 D_1 \sqrt{\frac{v_1}{v_2 v_3}} - 
D_2 \sqrt{\frac{v_2}{v_1 v_3}}  - D_3
\sqrt{\frac{v_3}{v_1 v_2}} \bigr)^2 \tau_{og} \nonumber \\
&&+ \bigl( N \sqrt{v_1v_2v_3} -
 D_1 \sqrt{\frac{v_1}{v_2 v_3}} + 
D_2 \sqrt{\frac{v_2}{v_1 v_3}}  - D_3
\sqrt{\frac{v_3}{v_1 v_2}} \bigr)^2 \tau_{of} \nonumber \\ 
& &+ \bigl( N \sqrt{v_1v_2v_3} -
 D_1 \sqrt{\frac{v_1}{v_2 v_3}} - 
D_2 \sqrt{\frac{v_2}{v_1 v_3}}  + D_3
\sqrt{\frac{v_3}{v_1 v_2}} \bigr)^2 \tau_{oh}
 \ \biggr\}
\label{s10}
\ea
displays four independent squared reflection coefficients, related to
the 9-branes and to the three types of 5-branes, and 
all terms enter this expression 
with signs fixed by the relation between
spin and statistics for the open spectrum in ${\cal A}$. 

These amplitudes determine by standard methods the vacuum channel
of the M\"obius amplitude  at the origin of the lattices
\ba
\tilde{\cal M}_0 &=& - \frac{1}{4} \biggl\{
\bigl[ N v_1v_2v_3 +
 D_1 \frac{v_1}{v_2 v_3} +  
D_2 \frac{v_2}{v_1 v_3}  + D_3
\frac{v_3}{v_1 v_2} \bigr] \hat{T}_{oo}  \nonumber \\
&+& \bigl[ (N + D_1 )v_1 + ( D_3 +  D_2)
\frac{1}{v_1} \bigr] \hat{T}_{og} +
\bigl[ (N + D_2 )v_2 + (D_3 + D_1)
\frac{1}{v_2} \bigr] \hat{T}_{of} \nonumber \\
&+&
\bigl[ (N + D_3 )v_3 + (D_2 + D_1)
\frac{1}{v_3} \bigr] \hat{T}_{oh} \biggr\} \quad .
\label{s11}
\ea
Here, this and all other M\"obius amplitudes are expressed in
terms of a convenient real basis of ``hatted'' characters. These may
defined starting from
\be
\chi(q) = q^{h - c/24} \sum_n \ d_n \ q^n \quad ,
\ee
letting $q\rightarrow e^{i \pi} q$ and removing an overall phase,
so that
\be
\hat{\chi}(q) = q^{h - c/24} \sum_n \ (-1)^n \ d_n \ q^n \quad .
\ee
For the hatted characters, the modular transformation connecting 
direct and transverse M\"obius channels is $P = T^{1/2} S T^2 S
T^{1/2}$, where $T: \tau \rightarrow \tau + 1$ and $S: \tau
\rightarrow - 1/\tau$.
The whole $\tilde{\cal M}$ is then
\ba
&&\!\!\!\!\!\!\!\!\!\!\!\!\!\!\!\tilde{\cal M} = - \frac{1}{4} \biggl\{
\bigl[ N v_1v_2v_3 W_1^e W_2^e W_3^e + 
\frac{ D_1 v_1}{v_2 v_3} W_1^e P_2^e P_3^e+ 
\frac{D_2 v_2}{v_1 v_3} P_1^e W_2^e P_3^e +  
\frac{D_3 v_3}{v_1 v_2} P_1^e P_2^e W_3^e\bigr] \hat{T}_{oo}  \nonumber \\
&+& \!\!\!\! \bigl[ ( N \!+\! D_1 )v_1 W_1^e \!+\! ( D_3 \!+\! D_2)
\frac{P_1^e}{v_1} \bigr] \hat{T}_{og} 
\left( \frac{2 \hat{\eta}}{\hat{\theta_2}}\right)^2 \!+\!
\bigl[ ( N \!+\! D_2 )v_2 W_2^e \!+\! ( D_3 \!+\! D_1)
\frac{P_2^e}{v_2} \bigr] \hat{T}_{of} 
\left( \frac{2 \hat{\eta}}{\hat{\theta_2}}\right)^2  \nonumber \\
&+&\!\!\!\!
\bigl[ ( N + D_3 )v_3 W_3^e + ( D_2 + D_1)
\frac{P_3^e}{v_3} \bigr] \hat{T}_{oh} 
\left( \frac{2 \hat{\eta}}{\hat{\theta_2}}\right)^2  \biggr\} \quad ,
\label{s12}
\ea
and a $P$ transformation recovers the direct-channel M\"obius
amplitude
\ba
&& {\cal M} = - \frac{1}{8} \biggl\{
\bigl[ N P_1 P_2 P_3 + 
 D_1 P_1 W_2 W_3+  
D_2 W_1 P_2 P_3 + D_3 W_1 W_2 P_3 \bigr] \hat{T}_{oo}  \nonumber \\
&-& \! \bigl[ (N \!+\! D_1 ) P_1 \!+\! (D_3 \!+\! D_2)
W_1 \bigr] \hat{T}_{og} 
\left( \frac{2 \hat{\eta}}{\hat{\theta_2}}\right)^2
\!-\! \bigl[ (N \!+\! D_2 )P_2 \!+\! (D_3 \!+\! D_1)
W_2 \bigr] \hat{T}_{of} 
\left( \frac{2 \hat{\eta}}{\hat{\theta_2}}\right)^2  \nonumber \\
&-&
\bigl[ (N + D_3 )P_3 + (D_2 + D_1)
W_3 \bigr] \hat{T}_{oh} 
\left( \frac{2 \hat{\eta}}{\hat{\theta_2}}\right)^2\biggr\} \quad . 
\label{s13}
\ea

The terms at the origin in (\ref{s7}), (\ref{s10})
and (\ref{s11}) combine, by construction, into the perfect
squares of the (untwisted) tadpole conditions
\be
N = D_1 = D_2 = D_3 = 32 \quad . \label{s14}
\ee
Notice that the M\"obius amplitude
reveals the presence of four symplectic gauge groups. The lack of
breaking terms, however, requires a rescaling of the four charges by a
factor two, in order to grant a proper particle interpretation, and the
resulting gauge group is $USp(16)^4$. This model was
discussed in \cite{bl} and, at a rational point, was previously 
considered in \cite{erice}. 

The massless spectrum can be easily read from ${\cal A}$ and
${\cal M}$. It is clearly not chiral, and is thus free of gauge
and gravitational anomalies. Rescaling the charges, so that $N=2n$ 
and $D_i=2 d_i$, one obtains
\ba
{\cal A}_0 &=& \frac{n^2 + d_1^2 +d_2^2 + d_3^2}{2} ( \tau_{oo} +
\tau_{og} + \tau_{oh} + \tau_{of} ) + (n d_1 + d_2 d_3 ) ( \tau_{go} +
\tau_{gg} + \tau_{gh} + \tau_{gf} )\nonumber \\
\!\!\!&+& \!\!\!
 (n d_2 + d_1 d_3 ) ( \tau_{fo} +
\tau_{fg} + \tau_{fh} + \tau_{ff} )+ (n d_3 + d_1 d_2 ) ( \tau_{ho} +
\tau_{hg} + \tau_{hh} + \tau_{hf} ) \ , \\
{\cal M}_0 &=&  \frac{1}{2}
(n + d_1 + d_2 + d_3)( \tau_{oo}-\tau_{og}-\tau_{of}-\tau_{oh}) \quad .
\ea
These amplitudes describe the
adjoint vector multiplets of the four $USp(16)$ groups, three chiral
multiplets, each  in the $(120,1,1,1)$,  $(1,120,1,1)$,  $(1,1,120,1)$ and
$(1,1,1,120)$, and six chiral multiplets in the $(16,16,1,1)$ and in
five additional bi-fundamental representations that differ by 
permutations of the factors.


\section{General properties of $Z_2 \times Z_2$ shift orientifolds}

If conventional orbifold operations are combined with shifts
$(\delta_L,\delta_R)$, the resulting
models provide string realizations \cite{kk} of the
Scherk-Schwarz mechanism \cite{ss}. 
In this Section we would like to describe the salient features of
the combined effects of shifts $(\delta_L,\delta_R)$
and $Z_2 \times Z_2$ orbifold operations on the open descendants of
type-IIB compactifications. As in \cite{ads,adds}, we
shall distinguish between symmetric {\it momentum} shifts 
$(p)=(\delta,\delta)$
and antisymmetric {\it winding} shifts 
$(w)=(\delta,-\delta)$, since the two  
have very different effects on the resulting spectra. 
These orbifolds correspond to singular limits of Calabi-Yau manifolds
with Hodge numbers $(19,19)$, $(11,11)$ and $(3,3)$ in the cases of
one, two and three shifts respectively.

Let us begin by introducing a convenient notation
to specify the orbifold action $X_i \to \sigma(X_i)$ on the
complex coordinates for the three internal tori. 
There are several ways to combine the three 
operations $g$, $f$ and $h$ of eq.  (\ref{s35})
with shifts in a way consistent with the $Z_2 \times Z_2$ group
structure. However, up to T-dualities and corresponding redefinitions
of the $\Omega$ projection, all non-trivial possibilities are captured by
\ba
\sigma_1(\delta_1,\delta_2,\delta_3) = \left( \begin{array}{rrrr}
\delta_1 & -\delta_2 & -1 \\
-1 & \delta_2 & -\delta_3 \\ -\delta_1 & -1 & \delta_3 \end{array}
\right) \quad , \qquad
\sigma_2(\delta_1,\delta_2,\delta_3) = \left( \begin{array}{rrrr} 
\delta_1 & -1 & -1 \\
-1 & \delta_2 & -\delta_3 \\ -\delta_1 & -\delta_2 & \delta_3 \end{array} 
\right)
\quad ,
\label{sigmas}
\ea
where the three lines refer to the new operations,
that we shall continue to denote by $g$, $h$ and $f$, and where
$-\delta_i$ indicates the combination of a shift in the
direction $i$ with an orbifold inversion. It is simple to convince
oneself that the two matrices $\sigma_1$
and $\sigma_2$ essentially 
exhaust all interesting possibilities within this class
of models. In order to reach this
conclusion, it is important to recognize
that when a shift only occurs in combination with an orbifold
inversion, it corresponds to a mere rotation of the fixed points, 
up to T-dualities and redefinitions of the $\Omega$ projection. 
Models of this type are {\it not} freely-acting orbifolds, but
rather conventional $Z_2 \times Z_2$ orbifolds with unconventional
$\Omega$ projections, and therefore are not discussed in this paper,
although in some cases they lead to interesting chiral spectra. For
instance, a similar winding shift eliminates some branes and yields chiral 
open descendants. Another
possibility, not considered here, is to combine the $\Omega$ projection 
with only one of the two group elements of $Z_2 \times Z_2$. 

T-duality transformations can be
neatly discussed referring to the tables (\ref{sigmas}), 
if these are supplemented by
an additional line corresponding to
the identity operation. A T-duality along a torus 
simply flips the signs in the corresponding column and interchanges
momentum and winding shifts. For instance, starting from $\sigma_1$
and performing a
T-duality along $T_{67}$ and $T_{89}$ would result in the table
\ba
\left( \begin{array}{rrrr} 1 & -1 & -1 \\ 
\delta_1 & \delta'_2 & 1 \\
-1 & -\delta'_2 & \delta'_3 \\ -\delta_1 & 1 & -\delta'_3 \end{array} \right)
\quad ,
\ea
where $\delta'$ denotes a dual shift. This is
equivalent to a standard table
$\sigma_1(\delta_1,\delta'_2,\delta'_3)$, up to a redefinition of
$\Omega$, that now includes a factor
$(-1)^{\delta_1+\delta'_2}$. 

In the following, we shall display a set
of independent models, corresponding to various choices of momentum
and winding shifts, while restricting our attention to 
a conventional $\Omega$ projection. 
Other choices for $\Omega$, including shifts, may be similarly discussed.
As in \cite{dp,abpss2}, they generally
affect the brane content, removing (or even adding) branes. With this
proviso, we can confine our attention to ten different types of 
$Z_2 \times Z_2$ shift orbifolds.

A simple and general rule predicts the types of allowed
branes, and will be justified in the following Sections:
\begin{quote}
{\it When a
line of the table contains $p$ or $-w$, the corresponding brane is
eliminated.}
\end{quote}

On the other hand, our choice of referring to a conventional 
$\Omega$ projection will always result in leaving the $D9$ branes
unaffected. Up to T-dualities and redefinitions of $\Omega$, one can
then shown that all interesting
brane configurations may be captured referring to the ten different
models of Table 1. 
\vskip 5pt
\begin{center}
\begin{tabular}{||c|c|c||}\hline\hline
{\rm 5-Branes}&{$\sigma_1$}&{$\sigma_2$} \\ \hline\hline
{0}&{$p_{123}$}&{$p_1w_2w_3$}\\
{}&{$w_{123}$}&{}\\ \hline
{1}&{}&{$w_1p_2p_3$} \\
{}&{$w_1p_2w_3$}&{}\\
{}&{}&{$w_1p_2$}\\
{}&{}&{$p_{23}$}\\ \hline
{2}&{$p_3$}&{$w_2p_3$} \\
{}&{}&{$w_1w_2p_3$}\\ \hline\hline
\end{tabular} 
\vskip 5pt
{Table 1. Types of shift models.}
\end{center}
\vskip 5pt

In listing these models, we have made suitable
choices of axes, so that when a single $D5$ brane is present, this is always
the first, $D5_1$, and when two are present, these are always
$D5_1$ and $D5_2$. All these models
have $N=1$ supersymmetry in the closed sector, but present interesting
instances of ``brane supersymmetry'' \cite{ads,kt}:  
additional 
supersymmetries are present 
for the massless modes, and at times also for the
massive ones \cite{bg}, confined to some branes. It is therefore 
interesting to display, for all the models of Table 1, the number of 
supersymmetries of the massless modes for the various branes present. 
The results are summarized in table 2:
\vskip 8pt
\begin{center}
\begin{tabular}{||c|c|c|c||}\hline\hline
{\rm models}&{$D9$ susy}&{$D5_{1}$ susy}&{$D5_{2}$ susy} \\ \hline\hline
{$p_3$}&{N=1}&{N=2} &{N=2}\\
{$w_2p_3$}&{N=2}&{N=2} &{N=4}\\ 
{$w_1w_2p_3$}&{N=4}&{N=4}&{N=4} \\  \hline
{$p_{23}$}&{N=1}&{N=2}&{--} \\
{$w_1p_2$}&{N=2}&{N=4}&{--} \\
{$w_1p_2p_3$}&{N=2}&{N=4}&{--} \\ 
{$w_1p_2w_3$}&{N=4}&{N=4}&{--} \\ \hline
{$p_{123}$}&{N=1}&{--}&{--} \\  
{$p_1w_2w_3$}&{N=2}&{--} &{--}\\ 
{$w_1w_2w_3$}&{N=4}&{--}&{--}
\\ \hline\hline
\end{tabular} 
\vskip 5pt
{Table 2. Brane supersymmetry for the various models.}
\end{center}
\vskip 8pt
Since the $p_{123}$ case was already discussed in \cite{adds},
in the following we can confine our attention to the nine remaining
classes of models. 

In all models with one or two shifts, the supersymmetry enhancement
for the massless modes on the branes may be traced directly to the
phenomenon described in \cite{ads,adds}: projections accompanied by
momentum shifts orthogonal to a brane do not reduce the supersymmetry
of the massless modes. As an example, let us consider the $w_1p_2$ model
that contains $D_9$ and $D5_1$ branes. After performing a T-duality
along the first and the third tori, the $D9$ branes turn into $D5_2$ 
branes and the $D5_1$ branes turn into $D5_3$ branes, while the resulting
model has momentum shifts in the first two directions. These are both
orthogonal to the $D5_3$ branes, and the corresponding massless modes
have indeed $N=4$ supersymmetry. On the other hand, one of these shifts
is parallel to the $D5_2$ branes and breaks the corresponding
supersymmetry to $N=2$. Models with three shifts are more subtle, due
to the $Z_2 \times Z_2$ group structure of the transformations, but
may be understood along similar lines.

With this proviso, we can conclude this Section by giving another 
simple rule, that will
be justified in the next Section, to predict the
amount of supersymmetry for the massless modes of a brane:
\begin{quote}
{\it Supersymmetry is enhanced for the massless modes on a
brane when the Scherk-Schwarz breaking is 
induced by momentum shifts orthogonal to it, or in all cases that may be linked to this by suitable
$T$-dualities.}
\end{quote}

\section{Structure of the shift orientifolds}

In this Section we would like to describe how to construct the vacuum
amplitudes for the open descendants of $Z_2 \times Z_2$ 
shift-orbifold  models.

In all cases the starting point is a deformation of the torus
amplitude (\ref{s1}), where momentum or
winding shifts induce corresponding modifications of the 
twisted contributions. For instance, for the case of two momentum shifts the
torus amplitude is
\ba
{\cal T}\!\!\!\!&=&\!\!\!{1 \over 4} \biggl\{ |T_{oo}|^2 \Lambda_1 \Lambda_2 \Lambda_3
\!+\!|T_{og}|^2 \Lambda_1 
|{4\eta^2 \over \theta_2^2}|^2 \!+\!
|T_{of}|^2 (-1)^{m_2} \Lambda_2 |{4\eta^2 \over \theta_2^2}|^2  
\!+\! |T_{oh}|^2 (-1)^{m_3} \Lambda_3 |{4\eta^2 \over \theta_2^2}|^2
 \nonumber \\
&+& |T_{go}|^2 \Lambda_1
|{4\eta^2 \over \theta_4^2}|^2+ |T_{gg}|^2 \Lambda_1
 |{4 \eta^2 \over \theta_3^2}|^2
+ |T_{fo}|^2 \Lambda_2^{n_2+1/2}
|{4\eta^2 \over \theta_4^2}|^2 \label{www} \\ 
&+& |T_{ff}|^2 
(-1)^{m_2} \Lambda_2^{n_2+1/2} |{4\eta^2 \over \theta_3^2}|^2 
+ |T_{ho}|^2 
\Lambda_3^{n_3+1/2} |{4 \eta^2 \over \theta_4^2}|^2 + 
|T_{hh}|^2 (-1)^{m_3} \Lambda_3^{n_3+1/2} |{4 \eta^2 \over \theta_3^2}|^2
\biggr\} \quad  . \nonumber
\ea
In all these deformed models, the independent orbit related to the 
discrete torsion is absent.

The corresponding open descendants are essentially determined by the
direct-channel Klein-bottle amplitude ${\cal K}$ and by the
transverse-channel annulus amplitude $\tilde{\cal A}$. Moreover,
the former, to which we now turn, is fully specified by the
corresponding table ($\sigma_1$ or $\sigma_2$) and by our choice of
working with a conventional $\Omega$ projection.

The Klein-bottle amplitudes are generically
affected by the shifts, that can make the transverse-channel
contributions massive. As in the supersymmetric case
discussed in Section 2, the direct-channel
amplitude, ${\cal K}$, includes four distinct contributions,
corresponding to $P_1P_2P_3$, $P_1W_2W_3$, $W_1P_2W_3$ and
$W_1W_2P_3$ where, as in previous Sections, $P_i$ and $W_i$
indicate lattice sums restricted to zero windings or zero momenta,
respectively. These four terms
reflect the $Z_2 \times Z_2$ structure of these models, 
and determine their content of $D9$, $D5_1$, $D5_2$ and $D5_3$ branes.
Whenever these terms are
accompanied in ${\cal K}$ by phases induced by the shifts, the 
corresponding vacuum-channel contributions are lifted in mass, and tadpole
conditions eliminate the corresponding branes. In order to
describe this effect in some detail, let us identify the two types
of projected lattice operators corresponding to the
shift matrices $\sigma_1$ and $\sigma_2$ of eq. (\ref{sigmas}) :
\ba
\label{v1}
\!\!\!\!\!\!&&V_1 = e^{i(p_{1L} x_{1L} + p_{2L} x_{2L} + p_{3L} x_{3L} + (L
\leftrightarrow R) )} + (-1)^{\delta_1 + \delta_2} \ 
e^{i(p_{1L} x_{1L} - p_{2L} x_{2L} - p_{3L} x_{3L} + (L
\leftrightarrow R) )}  \\
\!\!\!\!\!\!&&+ (-1)^{\delta_2 + \delta_3} \  
e^{i(- p_{1L} x_{1L} + p_{2L} x_{2L} - p_{3L} x_{3L} + (L
\leftrightarrow R) )}
+ (-1)^{\delta_1 + \delta_3} \ 
e^{i(- p_{1L} x_{1L} - p_{2L} x_{2L} + p_{3L} x_{3L} + (L
\leftrightarrow R))}  \ , \nonumber \\
\!\!\!\!\!\!&&V_2 = e^{i(p_{1L} x_{1L} + p_{2L} x_{2L} + p_{3L} x_{3L} + (L
\leftrightarrow R) )} + (-1)^{\delta_1} \ 
e^{i(p_{1L} x_{1L} - p_{2L} x_{2L} - p_{3L} x_{3L} + 
(L \leftrightarrow R) )}   \label{v2} \\
\!\!\!\!\!\!&&+ (-1)^{\delta_2+\delta_3} \ 
e^{i(- p_{1L} x_{1L} + p_{2L} x_{2L} - p_{3L} x_{3L} + (L
\leftrightarrow R) )}
+ (-1)^{\delta_1 + \delta_2 + \delta_3} \ 
e^{i(- p_{1L} x_{1L} - p_{2L} x_{2L} + p_{3L} x_{3L} + (L
\leftrightarrow R) )} \ ,  \nonumber
\ea
where $(-1)^\delta$ is $(-1)^m$ for a momentum shift and $(-1)^n$ for a
winding shift. These operators combine with diagonal untwisted oscillator
contributions, such as $|\tau_{oo}|^2$, and
can thus flow both in the Klein bottle ${\cal K}$ and in the
transverse-channel
annulus amplitude $\tilde{\cal A}$. As we have seen, the four types of 
contributions to ${\cal K}$, that correspond to 
$P_1P_2P_3$, $P_1W_2W_3$, $W_1P_2W_3$
and $W_1W_2P_3$, involve restricted lattice sums. They arise from
operators like those in eqs. (\ref{v1}) and (\ref{v2}),
specialized to the cases
of vanishing momenta or windings, that are therefore
eigenvectors of $\Omega$. Moreover, for the $P_1P_2P_3$
case the eigenvalue is always one, while in the other cases it is
simple to see from eq. (\ref{v1}) that the
eigenvalues are $(-1)^{\delta_1 + \delta_2}$, $(-1)^{\delta_2 +
\delta_3}$ and $(-1)^{\delta_2 + \delta_3}$ for the $\sigma_1$ case,
and  $(-1)^{\delta_1}$, $(-1)^{\delta_2+\delta_3}$ and 
$(-1)^{\delta_1 + \delta_2 +  \delta_3}$ 
for the $\sigma_2$ case. The end result
is that, aside from twisted contributions, the Klein bottle 
projections for the two cases may be written in the form
\be
{\cal K}_1=\frac{1}{8} \left\{ P_1P_2P_3 + (-1)^{\delta_1 + \delta_2} 
P_1W_2W_3 +
(-1)^{\delta_2 + \delta_3} W_1P_2W_3 + (-1)^{\delta_1 + \delta_3} W_1W_2P_3
\right\}T_{oo}  \quad ,
\label{k1}
\ee
and
\be
{\cal K}_2=\frac{1}{8} \left\{ P_1P_2P_3 + (-1)^{\delta_1} P_1W_2W_3 +
(-1)^{\delta_2+\delta_3} W_1P_2W_3 + 
(-1)^{\delta_1 + \delta_2+\delta_3} W_1W_2P_3
\right\} T_{oo} \quad .
\label{k2}
\ee
In order to describe the brane content of a given model, one can then
specialize these equations to the corresponding choice of (momentum or
winding) shifts. Whenever a phase has a nontrivial effect 
on a restricted lattice sum, the corresponding
transverse-channel contribution is lifted in mass, and the tadpole
conditions eliminate the corresponding brane. For instance, for the
$P_1W_2W_3$ sum this is the case for a momentum shift along the first torus,
and/or for winding shifts along the other two tori. In order to
discuss a concrete example, let us consider the $p_{23}$ model. In
this case the only phases present are $\delta_2=m_2$ and
$\delta_3=m_3$, while the corresponding multiplication table is
$\sigma_2$. Then, aside from twisted contributions
\be
{\cal K} =\frac{1}{8} \left\{ P_1P_2P_3 + P_1W_2W_3 +
(-1)^{m_2} W_1P_2W_3 + (-1)^{m_3} W_1W_2P_3
\right\} T_{oo} \quad ,
\ee
and consequently one is left only with $D9$ and $D5_1$ branes.

The twisted contributions to the Klein-bottle amplitude can then be
induced most conveniently, in all models, from the terms in 
$\tilde{\cal K}$ at the origin of the lattices, that they are to
complete into perfect squares. We shall see several examples of this 
procedure in the following Sections.

The other crucial ingredient of the construction, to which we now turn,
is the transverse-channel annulus amplitude $\tilde{\cal A}$. This
presents some subtleties, since the shift orbifold 
restricts the allowed untwisted modes. These 
constraints have a simple geometrical origin: 
in general, the fixed tori of one operation are paired by 
the others and form multiplets, whose members must accommodate
identical brane sets. As a result, branes are accompanied
by a number of images, and the restrictions on the form of $\tilde{\cal
A}$ determine their minimal configurations. This is 
reminiscent of what happens
in ordinary orbifolds when, by a continuous deformation, a brane is
moved off a fixed point, but in shift orbifolds the presence of
image branes is in general unavoidable. 

In order to prepare the grounds for the discussion of the shift
orbifolds, it is instructive to see how the
displacement of branes works in the $T^4/Z_2$ orbifold. This
phenomenon, first discussed in \cite{gp}, may be given a complete 
description in conformal field theory starting from the torus amplitude
\be
{\cal T} = \frac{1}{2} \biggl\{ \Lambda^4 | Q_o + Q_v |^2 +
| Q_o - Q_v |^2 | \frac{2 \eta}{\theta_2} |^4 
+ | Q_s + Q_c |^2 | \frac{2 \eta}{\theta_4} |^4 
+ | Q_s - Q_c |^2 | \frac{2 \eta}{\theta_3} |^4 \biggr\} \ ,
\ee
where $\Lambda$ denotes a one-dimensional toroidal lattice sum and, 
as usual, we have introduced the four supersymmetric
combinations of $SO(4)$ characters
\ba
Q_o &=& V_4 O_4 - C_4 C_4 \quad , \qquad Q_v = O_4 V_4 - S_4 S_4 \ , \nonumber
\\
Q_s &=& O_4 C_4 - S_4 O_4 \quad , \qquad Q_c = V_4 S_4 - C_4 V_4 \ . \label{g1}
\ea
The Klein-bottle amplitudes are then
\ba
{\cal K} &=& \frac{1}{4} \biggl\{ (Q_o + Q_v ) ( P^4 + W^4 ) + 
2 \times 16 (Q_s + Q_c)
\bigl( \frac{ \eta}{\theta_4} \bigr)^2 \biggr\} \quad , \nonumber \\
\tilde{\cal K} &=& \frac{2^5}{4} \biggl\{ (Q_o + Q_v ) 
( v (W^e)^4 + \frac{1}{v} (P^e)^4 ) + 2 (Q_o - Q_v)
\bigl( \frac{ 2 \eta}{\theta_2} \bigr)^2 \biggr\} \quad ,
\label{g2}
\ea
where $P^4$ and $W^4$ denote the restrictions of the lattice sum 
$\Lambda^4$ to zero windings and to zero momenta, respectively.

The simplest configuration, where all $D5$ branes are at the same fixed point
and no Wilson lines are introduced, results in the gauge group 
$U(16)_9 \times U(16)_5$. The familiar spectrum \cite{bs,gp} 
comprises two pairs of
hypermultiplets in the $(120,1)$ and $(1,120)$ from the $99$ and
$55$ sectors, and an additional $59$ hypermultiplet in the $(16,16)$.
\vskip 15pt
\input epsf \centerline{ \epsfbox{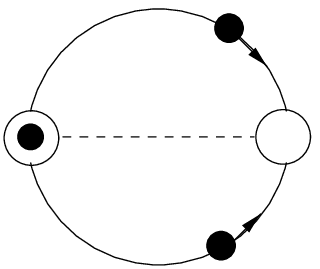}}
\begin{center}
Figure 1. Pairs of image branes moved away from a fixed point.
\end{center}

If some pairs of $D5$ branes are moved a
distance $2 \pi \alpha R$ away from the fixed point along the last
circle, the other transverse-channel amplitudes become
\ba
&& \tilde{\cal A} = \frac{2^{-5}}{4} \biggl \{ (Q_o+Q_v)\biggl[N^2 v W^4 
+ \frac{1}{v} \sum_m (D + \frac{\d}{2} e^{2i \p \a m}+\frac{\d}{2} e^{-2i \p \a m})^2P^3P_m
\biggr]  \nonumber \\
\!\!\!&+&\!\!\! 2N(D\!+\! \d )(Q_o\!-\!Q_v) \bigl( \frac{ 2 \eta}{\theta_2}
\bigr)^2  
\!+\! 4 (Q_s \!+\! Q_c ) \  (R_N^2 \!+\! R_D^2) \ 
\bigl( \frac{ 2 \eta}{\theta_4}
\bigr)^2 \!\!-\!\! 2 R_N R_D \ (Q_s \!-\! Q_c ) \bigl( \frac{2 \eta}{\theta_3}
\bigr)^2\biggr\}
 \ , \nonumber \\
\tilde{\cal M} &=&
- \frac{1}{2} \biggl \{ (\hat{Q}_o+\hat{Q}_v)\biggl[N v (W^{e})^4 
+ \frac{1}{v} 
\sum_m (D + \frac{\d}{2} e^{4i \p \a m}+\frac{\d}{2} e^{-4i \p \a
m}) (P^e)^3 P_{2m}
\biggr]  \nonumber \\
&+& (N+D+\d )(\hat{Q}_o-\hat{Q}_v)\left( 
\frac{\hat{\eta}}{\hat{\theta_2}}\right)^2 
\biggr\}
 \quad , \label{g3}
\ea
and thus the corresponding direct-channel amplitudes are
\ba
{\cal A} &\!\!=\!\!& \frac{1}{4} \biggl\{ (Q_o \!+\! Q_v ) \biggl [ N^2  P^4 
\!+\! (D^2 + \frac{\d^2}{2}) W^4 \!+\! \frac{\d^2}{4}W^3(W_{2 \a}+W_{-2\a})
\!+\! D \d W^3 (W_\a + W_{-\a}) \biggr]   \nonumber \\
\!\!\!&+&\!\!\! 2 N (D+\d ) (Q_s \!+\! Q_c ) \bigl(
\frac{\eta}{\theta_4} \bigr)^2
+  (Q_o \!-\! Q_v ) ( R_N^2 \!+\! R_D^2 ) \bigl( \frac{2 \eta}{\theta_2}
\bigr)^2 
+ 2 R_N R_D (Q_s \!-\! Q_c ) \bigl( \frac{\eta}{\theta_3} \bigr)^2
\biggr\} \ , \nonumber \\
{\cal M} &=& - \frac{1}{4} \biggl\{ (\hat{Q}_o + \hat{Q}_v ) 
\biggl[ N \ P^4 + D \ W^4 + \frac{\d}{2}W^3(W_{2 \a}+W_{-2\a}) \biggr]
\nonumber \\
&-& (\hat{Q}_o - \hat{Q}_v )(N + D + \d) 
\left( \frac{2\hat{\eta}}{\hat{\theta_2}} \right)^2
\biggr\} \quad .
\label{g4}
\ea
Here $N$ denotes the total number of $D9$ branes, 
$D$ denotes the total number of $D5$ branes left at the origin, 
$\delta$ counts the number of displaced branes, and $R_N$ and $R_D$
account for the effect of the orbifold breaking on the 
Chan-Paton charges at the fixed points. 
If $\alpha \neq 1/2$, the moved branes
are also away from the other fixed point on the circle, and
the corresponding breaking terms $R_{\d}$ are absent. Since
the consistency of the conformal theory in both channels allows only
breaking terms for branes sitting at fixed points,  
one is forced to let $\delta=2d$, thus
effectively splitting the charge between image branes. This is somewhat
reminiscent of the splitting induced by a quantized NS $B_{ab}$ \cite{Bab},
and indeed the rank of the gauge group is correspondingly reduced.
These features reflect
the geometry of the brane configuration, that 
includes pairs of images interchanged by orbifold operations.
The structure of ${\cal M}$ implies that in this case the gauge groups
carried by the displaced branes are symplectic and have a reduced
rank \cite{gp}. We would like to stress that,
while in the absence of breaking terms the $\delta$ contributions
appear to have enhanced supersymmetry, this is actually reduced 
by the M\"obius projection, as was the case for the supersymmetric
$Z_2 \times Z_2$ model without discrete torsion discussed in Section 2.
On the other hand, in the following Sections we shall see that 
in shift orbifolds 
${\cal M}$ respects the enhancement of supersymmetry.
Summarizing, for
$\alpha \neq 1/2$ the $D5$ gauge group generically breaks  
from $U(16)$ to $U(16-2d) \times USp(2d)$. On the other hand, for
$\alpha=1/2$ the pairs of images meet at the other fixed point on
the circle. The orbifold breakings are now allowed for all $D5$
branes, and the rank of the $D5$ gauge group is correspondingly
enhanced. In the transverse channel, the modified breaking
terms 
\be
\frac{2^{-5}}{4} \biggl \{ 
 16 (Q_s + Q_c ) \  (R_N^2 + R_D^2 + R_{\d}^2) \ 
\bigl( \frac{ \eta}{\theta_4}
\bigr)^2 \!\!-\!\! 8 R_N (R_D+R_{\d}) \ 
(Q_s \!-\! Q_c ) \bigl( \frac{ \eta}{\theta_3}
\bigr)^2\biggr\}  
\label{g5}
\ee
reflect the position of the $D5$ branes, that now occupy both
fixed points along the circle, and indeed at the origin eq. (\ref{g5})
becomes
\be
\frac{1}{4} \biggl[(R_N - 4R_{D})^2+(R_N - 4R_{\d})^2+14R_N^2 \biggr]
\quad .
\label{g6}
\ee

These results have a direct bearing on the main subject of this paper.
Indeed, if some orbifold operation were to interchange the two fixed
points where we have placed the $D5$ branes, as will be the case in
the $Z_2 \times Z_2$ shift orientifolds discussed in the following
Sections, one would be forced to set $D=\delta$ and $R_D=R_\delta$,
and as a result for $\alpha=1/2$ $\tilde{\cal A}$ would include the projector
\be
\Pi = \frac{1}{2} \left( 1 + (-1)^m \right) \quad .
\label{g7}
\ee
Summarizing, the presence of brane
multiplets in the vacuum configuration reflects itself in the 
presence of corresponding projection operators in $\tilde{\cal A}$.

We can now turn to describe how the projection operators may be
determined for the two classes of models corresponding to $\sigma_1$
and $\sigma_2$. The starting point are the projected lattice operators
$V_1$ and $V_2$ of eqs. (\ref{v1}) and (\ref{v2}), that combine
in the closed spectrum
with diagonal untwisted oscillator contributions, such as $\tau_{oo}$.
They are clearly 
allowed in $\tilde{\cal A}$, and determine the $NN$ and $D_iD_i$ terms,
and thus the brane content of the models.
If we let the corresponding states flow in the tube, all terms
in $V_1$ and $V_2$, being degenerate in mass, are accompanied by the
same power of $q$, and thus one inherits the
projectors
\ba
\Pi_1 &\sim& 1 + (-1)^{\delta_1+ \delta_2} + (-1)^{\delta_2+ \delta_3} + 
(-1)^{\delta_1+ \delta_3} \quad ,\\
\Pi_2 &\sim& 1 + (-1)^{\delta_1} + (-1)^{\delta_2+\delta_3} + 
(-1)^{\delta_1+ \delta_2+ \delta_3} \quad .
\label{g8}
\ea
In order to give a concrete example, let us refer again to
the $p_{23}$ model, that as we have seen has $D9$ and $D5_1$
branes. In this case
\be
\Pi = \frac{1}{2} \left( 1 + (-1)^{m_2+m_3} \right) \quad ,
\label{g9}
\ee
and all states flowing in the transverse channel of the annulus must
have correlated momenta, both even or both odd in the
corresponding two circles of the last two tori. Clearly, $\Pi$ has no effect
on the $D9$ contribution, that in
the transverse-channel annulus amplitude only involves
windings. However, it does play a crucial role for the sector
associated to the $D5_1$ brane. In this case, the transverse-channel
contribution corresponds to $W_1P_2P_3$, and the projection affects
the momenta in the last two sums. 
Since the Poisson transform of the restricted sum is proportional to
$P_1(W_2 W_3 +
W_2^{n+1/2} W_3^{n+1/2})$, where the $W_i^{n+1/2}$ in the last two
factors denote shifted sums, in the
direct channel $\Pi$  
gives rise to a doublet of branes. As we have seen, this is precisely
the type of configuration required by the structure of the shift orbifold.

\section{An interesting example with $N=2 \to N=1$ breaking}

In this Section we would like to describe in some detail an
interesting class of models, obtained introducing in 
the table  $\sigma_2$ of eq. (\ref{sigmas})  
momentum shifts $p_{2}$ and $p_{3}$ in two lattice directions,
where $N=2$ supersymmetry is
spontaneously broken to $N=1$. The previous analysis reveals that, in addition
to $D9$-branes, they also contain $D5_1$ branes. They have the
rather interesting feature of displaying various numbers of
supersymmetries in the bulk and on the branes in a relatively simple
setting.  The thumb rule 
of Section 3 predicts in this case $N=1$ supersymmetry for the $D9$
branes, since the two shifts are parallel to their world volume, and
$N=2$ supersymmetry for the massless modes of the $D5_1$ branes,
since they are acted upon by the $g$ projection, 
while the two shifts are orthogonal to their world volume.

The direct and transverse Klein bottle amplitudes for this model
\ba
{\cal K} &\!\!\!=\!\!\!&  \frac{1}{8} \biggl\{ T_{oo} \bigl[ P_1P_2P_3 \!+\!
P_1W_2W_3 \!+\! 
W_1(-1)^{m_2}P_2W_3 \!+\! W_1W_2(-1)^{m_3}P_3 \bigr] \!+\! 
2 \times 16 T_{go}\ P_1 
\left(\frac{ \eta}{\theta_4}\right)^2 \biggr\} , \nonumber \\
\tilde{\cal K} &\!\!\!=\!\!\!& \frac{2^5}{8} \biggl\{ 
T_{oo} \bigl[v_1v_2v_3W_1^eW_2^eW_3^e + 
\frac{v_1}{v_2v_3}W_1^eP_2^eP_3^e + \frac{v_2}{v_1v_3}P_1^eW_2^oP_3^e +
\frac{v_3}{v_1v_2}P_1^eP_2^eW_3^o\bigr] \nonumber \\
&\!\!\!+\!\!\!& 2T_{og}v_1W^e_1 \left(\frac{2 \eta}{\theta_2}\right)^2\biggr\} 
\quad ,
\label{e1}
\ea
are determined by the rules
described in Section 3. In particular, the twisted contribution is
fixed by the behavior of $\tilde{\cal K}$ at the origin of
the lattices, and ensures that the various 
independent sectors of the spectrum have reflection
coefficients that are perfect squares. Indeed:
\be
\tilde{\cal K}_0 = \frac{2^5}{8} \biggl\{ (\t_{oo}+\t_{og})
\biggl(\sqrt{v_1v_2v_3}+  \sqrt{\frac{v_1}{v_2v_3}}\biggr)^2 
+(\t_{of}+\t_{oh})\biggl(\sqrt{v_1v_2v_3}-  \sqrt{\frac{v_1}{v_2v_3}}
\biggr)^2 \biggr\} \ .
\label{e2}
\ee

The vacuum channel annulus amplitude
\ba
\tilde{\cal A} &=& \frac{2^{-5}}{8} \biggl \{
T_{oo}\biggl[N^2v_1v_2v_3W_1W_2W_3 
+ D_1^2\frac{v_1}{v_2v_3}W_1(P^{e}_2P^{e}_3 + P_2^oP_3^o) \biggr]  \nonumber \\
&+&  4(G^2 + 2 G_1^2) T_{go}W_1v_1 
\bigl( \frac{ 2 \eta}{\theta_4} \bigr)^2 + 
4F^2T_{fo}W_2^{n+1/2}v_2 \bigl( \frac{2 \eta}{\theta_4}\bigr)^2 
+ 4H^2T_{ho}W_3^{n+1/2}v_3 \bigl( \frac{2 \eta}{\theta_4}\bigr)^2 
\nonumber \\
&+& 2 N D_1 T_{og}W_1v_1\bigl( \frac{ 2 \eta}{\theta_2}\bigr)^2  
+ 4GG_1T_{gg}W_1v_1 \bigl( \frac{2 \eta}{\theta_3}\bigr)^2 
\biggr \}  
\label{e3}
\ea
inherits the projection operator
\be
\Pi= \frac{1}{2} \ [1+(-1)^{m_2+m_3}] 
\label{e4}
\ee
that, as we have seen in the previous Section, 
restricts the $55$ contribution to the even-even ($ee$) and 
odd-odd ($oo$) subsets
of momentum eigenvalues in the last two tori.  

Leaving a discussion of the breaking terms momentarily aside, 
by an $S$ transformation we can turn this
expression into the direct-channel annulus amplitude
\ba
{\cal A} &=& \frac{1}{8} \biggl\{ 
T_{oo} \biggl[ N^2P_1P_2P_3 + \frac{D_1^2}{2}P_1  (W_2W_3 \!+\! 
W_2^{n+1/2}W_3^{n+1/2}) \biggr] + (G^2 + 2 G_1^2) T_{og}P_1
\bigl( \frac{2 \eta}{\theta_2}\bigr)^2  
\nonumber \\
\!&+&\!  
F^2 \ T_{of}(-1)^{m_2}P_2\bigl( \frac{2 \eta}{\theta_2}
\bigr)^2 
+ H^2 \ T_{oh}(-1)^{m_3}P_3\bigl( \frac{2 \eta}{\theta_2}\bigr)^2   
+ 2 N D_1 T_{go}P_1\bigl( \frac{ \eta}{\theta_4}\bigr)^2  \nonumber \\
&+& 4GG_1T_{gg}P_1\bigl( \frac{ \eta}{\theta_3}\bigr)^2  \biggr\} \quad .
\label{e5}
\ea
Whereas the $99$ strings are rather conventional, and have the
usual three
projection terms, although affected by the momentum shifts as in
the simpler models of \cite{ads,adds}, the 
configuration of
the $55$ strings is more peculiar, and 
admits a nice geometrical interpretation: it corresponds
to a doublet of branes, associated to a pair of tori fixed by
$g$ and interchanged by $f$ and $h$. Only the 
$g$ projection is present, since in
this sector the physical states are combinations of pairs 
localized on the image branes. In this case the {\it full} 
$D5_1$ spectrum, not
only its massless modes, has enhanced supersymmetry. 
We have already stressed that multiplets of
branes are a generic feature of these shift orbifolds. Even
if one attempts to insert all branes at a fixed point, the other
operations typically move them, and give rise to
multiple images.
This is summarized for this class of models in figure 2, where
the three axes refer to the three two-tori $T_{45}$, $T_{67}$ and $T_{89}$.
As usual, the massive untwisted excitations of these orbifold
models have enhanced supersymmetry,  although they have 
only a fraction of the lattice modes present in the toroidal case. 
More precisely, the 
$g$ projection forbids this extension in the first torus, while in the
second and third tori the projected combinations of lattice states
have indeed $N=2$ supersymmetry in the $99$ sector and $N=4$ supersymmetry in
the $55$ sector.
\vskip 5pt
\input epsf \centerline{ \epsfbox{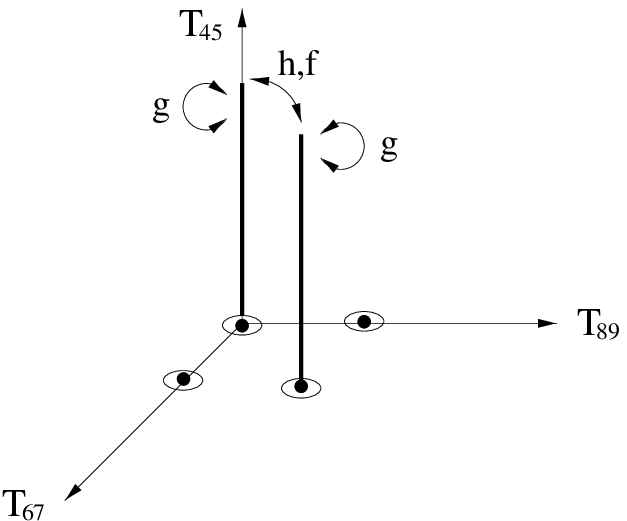}}
\begin{center}
Figure 2. $D5_1$ brane configuration for the $p_{23}$ model
\end{center}
\vskip 5pt
  
The previous considerations determine the structure of
the breaking terms in the transverse channel: as we have anticipated,
the $D5_1$ branes occupy a pair of fixed tori, while all others are empty. 
Then, according to the  rule discussed in the previous Section, in the
transverse
channel the breaking terms are to group into the structures
\be
\frac{2}{4}(G \mp 4G_1)^2 + \frac{14}{4}G^2 \ ,
\label{e6}
\ee
precisely as demanded by the direct-channel amplitude ${\cal A}$. The relative 
factor between $G$ and $G_1$ in the first term counts the number of occupied 
fixed points; more precisely in all these models it is equal to
\be
\sqrt{\frac{{\rm \#  \ of \ fixed \ points}}
{{\rm  \# \ of \ occupied \ fixed \ points}}} \quad .
\label{rule}
\ee
At the origin of the lattices, the untwisted contribution to the
transverse-channel annulus amplitude is
\be
\tilde{\cal A}_0 \!=\! \frac{2^{-5}}{8} \biggl\{ (\t_{oo}\!+\!
\t_{og})\biggl(N\sqrt{v_1v_2v_3}\!+\! D_1 \sqrt{\frac{v_1}{v_2v_3}}\biggr)^2 
\!+\!(\t_{of}\!+\!\t_{oh})\biggl(N\sqrt{v_1v_2v_3}\!-\! 
D_1 \sqrt{\frac{v_1}{v_2v_3}}\biggr)^2  \biggr\} \ ,
\label{e7}
\ee
and, together with ${\tilde{\cal K}}_0$ of eq. (\ref{e2}), determines the
transverse-channel M\"obius amplitude 
\ba
\tilde{\cal M} &=& - \frac{1}{4} \biggl\{ 
\hat{T}_{oo}\biggl[Nv_1v_2v_3W_1^eW^e_2W_3^e  
+ D_1 \frac{v_1}{v_2v_3} W_1^eP_2^{e}P_3^{e}\biggr] 
+ \hat{T}_{og}
(N+D_1)W^e_1v_1\left( \frac{2 \hat{\eta}}{\hat{\theta_2}}\right)^2  
 \nonumber \\
&+& N\hat{T}_{of}W_2^ov_2 \left( \frac{2 \hat{\eta}}{\hat{\theta_2}}\right)^2  
- N\hat{T}_{oh}W_3^ov_3 \left( \frac{2 \hat{\eta}}{\hat{\theta_2}}\right)^2  
\biggr\} \quad .
\label{e8}
\ea
Finally, a $P$ transformation determines the direct-channel M\"obius
projection
\ba
{\cal M} &\!\!=\!\!& - \frac{1}{8} \biggl\{ 
\hat{T}_{oo} \biggl[NP_1P_2P_3 \!+\!  
D_1 P_1W_2W_3 \biggr]
\!-\! \hat{T}_{og}(D_1+ N)P_1 \left( \frac{2 \hat{\eta}}{\hat{\theta_2}}
\right)^2 \!-\! N\hat{T}_{of}(-1)^{m_2}P_2
\left( \frac{2 \hat{\eta}}{\hat{\theta_2}}\right)^2    \nonumber \\
&+& N\hat{T}_{oh}(-1)^{m_3}
P_3 \left( \frac{2 \hat{\eta}}{\hat{\theta_2}}\right)^2  
 \biggr\} \quad ,
\label{e9}
\ea
whose contributions at the origin of the lattices imply that both
types of gauge groups are unitary. The untwisted tadpole conditions
\be
N = D_1 = 32
\label{e10}
\ee
fix the (maximal) size of the two gauge groups, while the twisted
tadpole conditions
\be
G = G_1 = 0 
\label{e11}
\ee
are identically satisfied if one parametrizes the charges according to
\ba
N &=& o+g+\bar{o}+\bar{g} \ , \hspace{1cm}
G = i(o+g-\bar{o}-\bar{g}) \ , \nonumber \\
H &=& (o-g+\bar{o}-\bar{g}) \ , \hspace{1cm}
F=i(o-g-\bar{o}+ \bar{g}) \ , \nonumber \\
D_1 &=& 2(d+\bar{d}) \ , \hspace{2cm} G_1= i(d - \bar{d}) \ ,
\label{e12}
\ea
provided $o+g=16$ and $d=8$.
The massless modes from the open sector are then summarized in
\ba
{\cal A}_0 &+&{\cal M}_0= \ \bigl( o\bar{o} + g\bar{g} \bigr)
\t_{oo} + \bigl( o \bar{g}+ g \bar{o} \bigr) \t_{og} + \bigl( o g +
\bar{o}\bar{g} \bigr)\t_{of} \nonumber \\
&+& \biggl[ \frac{o(o-1)}{2}+\frac{g(g-1)}{2}+
\frac{\bar{o}(\bar{o}-1)}{2}+\frac{\bar{g}(\bar{g}-1)}{2} \biggr] \t_{oh}
 \nonumber \\ 
&+& d \bar{d} ( \t_{oo}+\t_{og} )
+ \biggl[ \frac{d(d-1)}{2}+\frac{\bar{d}(\bar{d}-1)}{2} 
\biggr](\t_{of}+\t_{oh})
 \label{e13} \\
&+&  \bigl[ (o+g)d+(\bar{o}+\bar{g})\bar{d} \bigr](\t_{gf}+\t_{gh})
\nonumber  \quad ,
\ea
while
the gauge groups allowed by the tadpole conditions in this class of
models are
\be
[U(o)\times U(g)]_9 \times U(8)_{5_1} \ .
\label{e14}
\ee

The tree-level spectrum of the $D9$ branes has $N=1$ supersymmetry,
while the whole spectrum of the $D5_1$ branes, orthogonal to the 
directions used to induce
supersymmetry breaking, has indeed $N=2$ supersymmetry.  This is one more
instance of the phenomenon discussed in \cite{ads,adds}, and extended
to the whole spectrum in \cite{bg}, as suggested in \cite{kt}. The massless
spectrum is not chiral, and includes $99$ chiral multiplets in the
representations $(o(o-1)/2,1,1)$, $(1,g(g-1)/2,1)$, 
$(o,\bar{g},1)$, $(o,g,1)$ and their hermitian conjugates, 
one pair of $55$ hypermultiplets in the representation $(1,1,d(d-1)/2)$, and
$59$ hypermultiplets in the representations 
$(o,1,d)$ and $(1,g,d)$.


It is instructive to discuss the behavior of this spectrum in the
decompactification limit for the directions used to break supersymmetry. 
In \cite{adds} we studied some simpler examples, where momentum
(or winding) shifts along one coordinate were combined with a single 
$Z_2$ orbifold inversion in other directions. In
these cases it is relatively
simple to understand how, in the singular $R \to \infty$ ($R \to
0$) limit, the deformed spectrum collapses into a continuum of
momentum (winding) modes with N=4 supersymmetry. On the other hand, in
the decompactification limit, the
$p_2p_3$ models discussed in this section recover only N=2
supersymmetry. This is already a subtler setting, even in the
supersymmetric case without shifts: in this limit
the untwisted bulk states are still in the presence of walls, so that
their momentum  modes are still projected, while some twisted states 
are moved to infinity, together with the corresponding fixed
points. In these shifted non-compact orbifolds, the momentum modes for the
untwisted states merge in the limit in an extended continuum, but some
of the
fixed points, with the corresponding brane content, move
away to infinity. Moreover, in these open-string models the singular
limit is usually accompanied by the emergence of new tadpoles
\cite{pw,ads}, that may be eliminated arranging for {\it local}
cancellations \cite{ab}. Actually, in this case if only one of the radii $R_2$
and $R_3$ tends to infinity, the local tadpole conditions are
identically
satisfied. This fact has a simple geometric reason: the image doublets
of $D5_1$ branes saturate locally the RR charge. We would
like to stress that
the identified images carry the same gauge group, a fact to be
contrasted with the simpler toroidal setting of \cite{pw}.
On the
other hand, when both radii are large, the saturation requires 
a further breaking of the $D5_1$ gauge 
group to $U(4) \times U(4)$.
In the decompactification limit of these $p_2p_3$
models, a single set of $D5$ branes is left, with a
$U(4)$ gauge group, while the
others are moved 
to infinity. This limiting configuration
 may still be linked to the conventional $U(16)_9 \times
U(16)_5$ setting of the compact $T^4/Z_2$ orbifold, albeit 
in a $(U(4)_5)^4$ configuration, and where 12 of the $D5_1$
branes have been moved to infinity in the decompactification limit.

\section{Freely-acting orbifold models with two $D5$ branes}

In this Section we describe other classes of
shift models of this type, with two $D5$ branes. Our aim is to give
a geometrical
interpretation of the corresponding brane multiplets and to exhibit
their $M$ theory limits, whenever they exist, as well as their
massless spectra. For the sake of
brevity, we shall only display direct-channel amplitudes.

\subsection{ One momentum shift}

This case corresponds to introducing a single momentum shift $p_3$ in 
the table $\sigma_1$ of eq. (\ref{sigmas}). The resulting models 
contain $D9$, $D5_1$ and $D5_2$ branes.
The direct-channel Klein bottle amplitude is
\ba
{\cal K} &=& \frac{1}{8} \biggl\{ T_{oo}\biggl[P_1P_2P_3 + P_1W_2W_3 
+ W_1P_2W_3 + W_1W_2(-1)^{m_3}P_3\biggr]
\nonumber \\
&+& 2 \times 16 T_{go}P_1\left(\frac{ \eta}{\theta_4}\right)^2 
+ 2 \times 16 T_{fo}P_2\left(\frac{ \eta}{\theta_4}\right)^2 
+ 2 \times 16 T_{ho}W_{3}^{n+1/2}\left(\frac{ \eta}{\theta_4}\right)^2 
\biggr\} \quad ,
\label{d1}
\ea
the direct-channel annulus amplitude is
\ba
{\cal A} &=& \frac{1}{8} \biggl\{ T_{oo}\biggl[
N^2 P_1P_2P_3 + \frac{D_1^2}{2} P_1W_2(W_3+W_3^{n+1/2})  
 + \frac{D_2^2}{2} W_1P_2(W_3+W_3^{n+1/2})  \biggr]  \nonumber \\
&+&  (G^2 + 2 G_1^2) T_{og}P_1\left( \frac{2\eta}{\theta_2} \right)^2 
+ (F^2 + 2 F_2^2) T_{of} P_2 \left( \frac{2\eta}{\theta_2} \right)^2
+ H^2T_{oh}(-1)^{m_3}P_3\left( \frac{2\eta}{\theta_2} \right)^2 \nonumber \\
&+& 2ND_1T_{go}P_1\left( \frac{\eta}{\theta_4} \right)^2 
+ 2ND_2 T_{fo}P_2\left( \frac{\eta}{\theta_4} \right)^2  
+D_1D_2T_{ho}(W^{n+1/4}_3\!+\!W_3^{n+3/4}) 
\left( \frac{\eta}{\theta_4} \right)^2 \nonumber \\
&+& 4GG_1T_{gg}P_1\left( \frac{\eta}{\theta_3} \right)^2 
\!+\! 4FF_2T_{ff}P_2\left( \frac{\eta}{\theta_3} \right)^2 \biggr\} \ ,  
\label{d2}
\ea
and finally the M\"obius projection is
\ba
\!\!\!\!\!\!\!\!\!&&{\cal M} = - \frac{1}{8} \biggl\{ 
\hat{T}_{oo}\biggr[NP_1P_2P_3 
+D_1P_1W_2W_3 + D_2W_1P_2W_3\biggr]
- \hat{T}_{og}(D_1+N)P_1\left( \frac{2 \hat{\eta}}{\hat{\theta_2}} \right)^2  
\nonumber \\ \!\!\!\!\!\!\!\!\!\!&-&\!\!\!\! \hat{T}_{of}(D_2+N)P_2 
\left( \frac{2 \hat{\eta}}{\hat{\theta_2}} \right)^2 
\!\!-\! \hat{T}_{oh}(D_2+D_1)W_3^{n+1/2}
\left( \frac{2 \hat{\eta}}{\hat{\theta_2}}\right)^2 
\!\!+\! \hat{T}_{oh}N(-1)^{m_3}P_3
\left( \frac{2 \hat{\eta}}{\hat{\theta_2}}\right)^2  
\biggr\} \ .
\label{d3}
\ea
The corresponding brane configuration is shown in Figure 3, where the
$D5_1$ branes are parallel to the $T_{45}$ axis and the $D5_2$ branes
are parallel to the $T_{67}$ axis.
\vskip 5pt
\input epsf \centerline{ \epsfbox{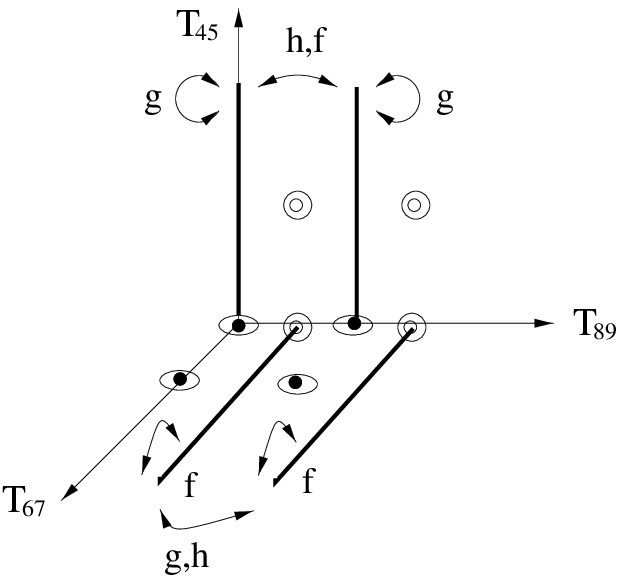}}
\begin{center}
Figure 3. $D5_1$ and $D5_2$ brane configuration for the $p_{3}$ model
\end{center}
\vskip 5pt

We can parameterize the charges according to
\ba
N &=& o+g+\bar{g}+\bar{o} \ , \hspace{1cm} 
G = i(o+g-\bar{g}-\bar{o}) \ ,  \nonumber \\
H &=& o-g-\bar{g}+\bar{o} \ , \hspace{1cm} 
F = i(o-g+\bar{g}-\bar{o}) \ , \nonumber \\
D_1 &=& 2(d_1 + \bar{d}_1) \ , \hspace{1cm} G_1 = i(d_1 - \bar{d}_1) \ ,
\nonumber \\
D_2 &=& 2(d_2 + \bar{d}_2) \ , \hspace{1cm} F_2 = i(d_2 - \bar{d}_2) \ ,
\label{d4}
\ea
and the tadpole conditions 
\be
o+g = 16 \ , \hspace{2cm} d_1 = 8 \ , \hspace{2cm}
d_2 = 8  \ 
\label{d5}
\ee
then require a gauge group
\be
[U(o) \times U(g)]_{9} \times [U(8)]_{5_1} \times [U(8)]_{5_2} \ .
\label{d6}
\ee
The massless spectrum is again not chiral and has $N=1$ supersymmetry
on the $D9$ branes and $N=2$ supersymmetry on the two types 
of $D5$ branes, that are
orthogonal to the direction used for the breaking. 
The $99$ spectrum contains chiral multiplets in the 
representations $(o(o-1)/2,1,1,1)$,
$(1,g(g-1)/2,1,1)$, $(o,g,1,1)$ and $(o,\bar{g},1,1)$ and their
conjugates.  On the other hand,
the $55$ spectrum contains pairs of $N=2$ hypermultiplets in the
representations $(1,1,d_1(d_1-1)/2,1)$ and  
$(1,1,1,d_2(d_2-1)/2)$. Finally, the $59$ spectrum includes
hypermultiplets in the representations $(o,1,d_1,1)$, $(1,g,d_1,1)$,
$(o,1,1,d_2)$ and $(1,\bar{g},1,d_2)$.

The brane content of the $p_2p_3$ models discussed in the previous 
section is effectively
a truncation of this, since the additional momentum shift eliminates
the $D5_2$ branes. Here one can follow  more closely 
the decompactification limit
of the single direction that accommodates the shift and, 
interestingly, the resulting low-lying modes originate from a
superposition of states in two orbifold directions. Thus, the two
orthogonal sets of $D5$ branes are effectively {\it parallel} in this limit,
where a more conventional $U(16)_9 \times (U(8)_5)^2$ structure is
recovered. The annulus amplitude is compatible from the start with
local tadpole cancellations, because of the doublet structure of the
$D5_1$ and $D5_2$ branes. Moreover, two sets of $D5$ branes are
present in the limit, where the corresponding $U(8) \times U(8)$ gauge
group finds
the usual explanation in terms of Horava-Witten walls \cite{hw}, as in the 
simpler examples of \cite{ads,adds}.


\subsection{One winding shift and one momentum shift}

This case corresponds to introducing a winding shift $w_2$ and a
momentum shift $p_3$ in 
the table $\sigma_2$ of eq. (\ref{sigmas}). The resulting models 
contain $D9$, $D5_1$ and $D5_2$ branes.
The direct-channel Klein bottle amplitude is
\ba
{\cal K} &=& \frac{1}{8} \biggl\{ T_{oo}\biggl[P_1P_2P_3 
+ P_1W_2W_3 + W_1P_2W_3 
+ W_1(-1)^{n_2}W_2(-1)^{m_3}P_3 \biggr]
\nonumber \\
&+& 2 \times 16 T_{go}P_1\left(\frac{ \eta}{\theta_4}\right)^2 
+ 2 \times 16 T_{fo}P^{m+1/2}_2\left(\frac{ \eta}{\theta_4}\right)^2 
+ 2 \times 16 T_{ho}W_3^{n+1/2}\left(\frac{ \eta}{\theta_4}\right)^2 
\biggr\} \quad ,
\label{d7}
\ea
the direct-channel annulus amplitude is
\ba
{\cal A} &=& \frac{1}{8} \biggl\{ 
T_{oo}\biggl[\frac{N^2}{2}P_1(P_2+P_2^{m+1/2})P_3 
+\frac{D_1^2}{2}P_1W_2(W_3+W_3^{n+1/2})  \nonumber \\
&+& \frac{D_2^2}{4}W_1(P_2+P_2^{m+1/2})(W_3+W_3^{n+1/2}) \biggr]
+2(G^2+G_1^2)T_{og}P_1\left( \frac{2\eta}{\theta_2} \right)^2 \nonumber \\
&+& 2ND_1T_{go}P_1\left( \frac{\eta}{\theta_4} \right)^2 
+ ND_2T_{fo}(P_2^{m+1/4}+P_2^{m+3/4})\left( \frac{\eta}{\theta_4} \right)^2 
\nonumber \\
&+& D_1D_2T_{ho}(W_3^{n+1/4}+W_3^{n+3/4})
\left( \frac{\eta}{\theta_4} \right)^2 
+ 8GG_1T_{gg}P_1 \left( \frac{\eta}{\theta_3} \right)^2
\biggr\}
\quad ,
\label{d8}
\ea
and finally the M\"obius projection is
\ba
{\cal M} &=& - \frac{1}{8} \biggl\{ \hat{T}_{oo}\biggl[NP_1P_2P_3 
+D_1P_1W_2W_3  + D_2W_1P_2W_3\biggr] - \hat{T}_{og}(D_1+N)P_1
\left( \frac{2 \hat{\eta}}{\hat{\theta_2}}\right)^2  
\nonumber \\
&-& \hat{T}_{of}(D_2+N)P_2^{m+1/2}\left( \frac{2 \hat{\eta}}{\hat{\theta_2}}
\right)^2  - \hat{T}_{oh}(D_2+D_1)W_3^{n+1/2}
\left( \frac{2 \hat{\eta}}{\hat{\theta_2}} \right)^2    \biggr\} \quad .
\label{d9}
\ea
\vskip 5pt
\input epsf \centerline{ \epsfbox{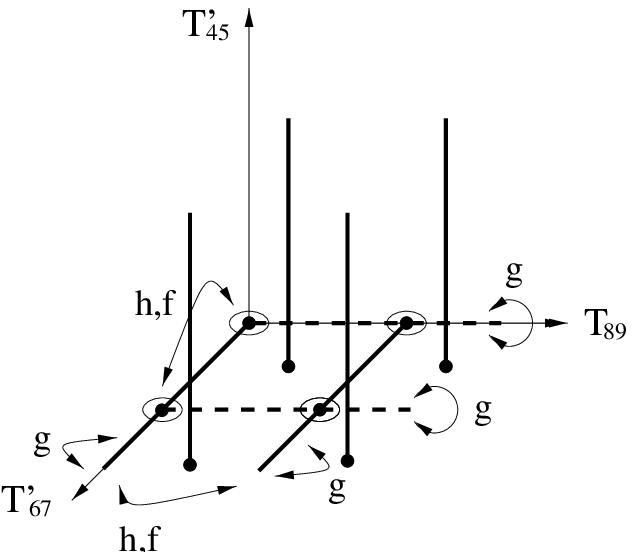}}
\begin{center}
Figure 4. $D_9$ (dashed), $D5_1$ and $D5_2$ branes for
the $w_2p_{3}$ model.
\end{center}
\vskip 5pt

Figure 4 shows the brane configuration, after a T-duality along $T_{45}$
and $T_{67}$, that turns the $D9$ branes into $D5'_3$ branes (dashed), 
the $D5_1$ branes
into $D5'_2$ branes (horizontal) and the $D5_2$ branes into $D5'_1$
branes (vertical). This model may actually be related to the
$p_2p_3$ model discussed in Section 5, via a T-duality in the first
two tori. Still, we have chosen to display it, since it shows how a
simple redefinition of $\Omega$ can affect both the brane distribution
and the superymmetry of the massless modes.

We can parameterize the charges according to
\ba
N &=& 2(n+\bar{n}) \ , \hspace{1cm}  D_1 = 2(d_1+\bar{d}_1) \ , 
\hspace{1cm} D_2 = 4d_2 \ , 
\nonumber \\ 
G &=& i(n-\bar{n}) \ , \hspace{1cm}  G_1 = i(d_1 - \bar{d}_1) \ ,
\label{d10}
\ea
and the tadpole conditions 
\be
n = 8 \ , \hspace{2cm}
d_1 = 8 \ , \hspace{2cm}  d_2 = 8 
\label{d11}
\ee
then require a gauge group
\be
U(8)_9 \times  U(8)_{5_1} \times  SO(8)_{5_2} \quad .
\label{d12}
\ee

The massless spectrum is again not chiral and has $N=2$ supersymmetry
on the $D9$ and $D5_1$ branes, and $N=4$ supersymmetry on the
$D5_2$ branes. Our thumb rule explains rather naturally the first two
results: the $D9$ branes are not affected by the (parallel) winding shift, 
while the $D5_1$ branes are not affected by the (orthogonal) momentum shift.
 
Aside from the gauge multiplets, the massless
spectrum contains pairs of $N=2$ hypermultiplets in the
representation $(28,1,1)$ from the 
$99$ sector, in the representation $(1,28,1)$ from the 
$5_15_1$ sector, and in the representation $(8,8,1)$ from the $95_1$
sector.

In this case the interesting limits  are $R_2 \to 0$ and $R_3 \to
\infty$. Now {\it all} local tadpoles are canceled, even in the
simultaneous limit, without the need of any further splitting. 
This interesting property finds again a natural
explanation in the geometry of the brane multiplets. Indeed, while
the $D9$ and $D5_1$ branes, orthogonal to one of the directions
where shifts have been introduced, are properly arranged in
doublets, the $D5_2$ branes, orthogonal to both, are arranged in 
quadruplets. This structure is, again, exactly as needed to ensure
local tadpole cancellation. Moreover, as quadruplets are moved
by all projections, their massless modes have $N=4$ supersymmetry.
\subsection{Two  winding shifts and one momentum shift}

This case corresponds to introducing a momentum shift $p_3$ along the
third torus, and winding shifts $w_1$ and $w_2$ along the others in
the table $\sigma_2$ of eq. (\ref{sigmas}).
The resulting models contain $D9$, $D5_1$ and $D5_2$ branes.
The direct-channel Klein bottle amplitude is
\ba
{\cal K} &=& \frac{1}{8} \biggl\{ T_{oo}\biggl[P_1P_2P_3 
+ P_1W_2W_3 + W_1P_2W_3+ (-1)^{n_1}W_1(-1)^{n_2}W_2(-1)^{m_3}P_3 \biggr]
\nonumber \\
\!\!\!\!&+&\!\!\!\! 2 \times 16 T_{go}P^{m+1/2}_1\left(\frac{ \eta}{\theta_4}
\right)^2 
\!\!\!+\!\! 2 \times 16 T_{fo}P^{m+1/2}_2\left(\frac{ \eta}{\theta_4}\right)^2 
\!\!\!+\! 2 \times 16 T_{ho}W_3^{n+1/2}\left(\frac{ \eta}{\theta_4}\right)^2 
\biggr\} \ ,
\label{d15}
\ea
the direct-channel annulus amplitude is
\ba
{\cal A} &=& \frac{1}{8} \biggl\{ 
T_{oo}\biggl[\frac{N^2}{4}(P_1+P_1^{m+1/2})(P_2+P_2^{m+1/2})P_3
+ \frac{D_1^2}{4}(P_1+P_1^{m+1/2})W_2(W_3+W_3^{n+1/2}) \nonumber \\
&+& \frac{D_2^2}{4}W_1(P_2+P_2^{m+1/2})(W_3+W_3^{n+1/2}) \biggr] 
+T_{go}ND_1(P_1^{m+1/4}+P_1^{m+3/4})\left( \frac{\eta}{\theta_4} \right)^2  
\nonumber \\
\!\!\!&+&\!\!\! 
T_{fo}ND_2(P_2^{m+1/4}+P_2^{m+3/4})\left( \frac{\eta}{\theta_4} \right)^2
+T_{ho}D_1D_2(W_3^{n+1/4}+W_3^{n+3/4})\left( \frac{\eta}{\theta_4} \right)^2
] \biggr\}  \ ,
\label{d16}
\ea
and finally the M\"obius projection is
\ba
{\cal M} &=& - \frac{1}{8} \biggl 
\{ \hat{T}_{oo} \biggl[ NP_1P_2P_3 + D_1P_1W_2W_3 + D_2W_1P_2W_3 \biggr]
- \hat{T}_{og}
(D_1+N)P_1^{m+1/2}\left( \frac{2 \hat{\eta}}{\hat{\theta_2}}\right)^2  
\nonumber \\
&-& \hat{T}_{of}
(D_2+N)P_2^{m+1/2}\left( \frac{2 \hat{\eta}}{\hat{\theta_2}}\right)^2 
- \hat{T}_{oh}(D_1+D_2)W_3^{n+1/2} \left( \frac{2 \hat{\eta}}{\hat{\theta_2}}
\right)^2  \biggr\} \ .
\label{d17}
\ea
\vskip 5pt
\input epsf \centerline{ \epsfbox{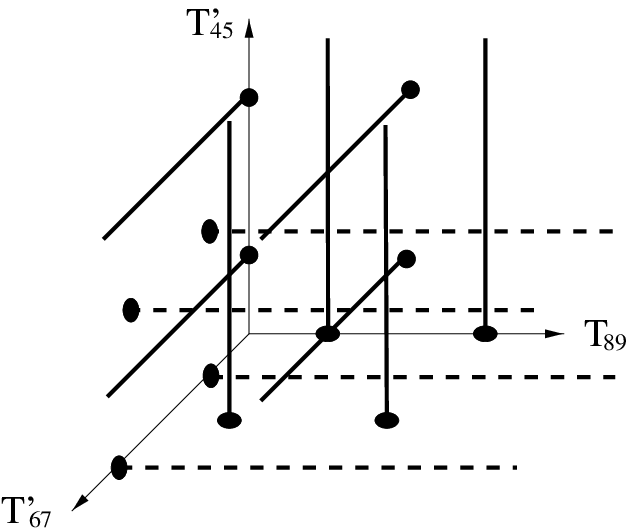}}
\begin{center}
Figure 5. $D_9$ (dashed), $D5_1$ and $D5_2$ branes for
the $w_1w_2p_3$ model.
\end{center}
\vskip 5pt

Figure 5 shows the brane configuration, after a T-duality along $T_{45}$
and $T_{67}$, that turns the $D9$ branes into $D5'_3$ branes (dashed), 
the $D5_1$ branes
into $D5'_2$ branes (horizontal) and the $D5_2$ branes into $D5'_1$
branes (vertical).

We can parameterize the charges according to
\be
N = 4n \ , \hspace{1cm} D_1 = 4d_1 \ , \hspace{1cm} D_2 = 4d_2 \quad ,
\label{d18}
\ee
and the tadpole conditions
\be
n = d_1 = d_2 = 8 \ 
\label{d19}
\ee
then require a gauge group
\be
SO(8)_9 \times  SO(8)_{5_1} \times  SO(8)_{5_2} \quad .
\label{d20}
\ee

The massless spectrum is again not chiral, but now has $N=4$ supersymmetry
on all the branes. The interesting limits are in this case
$R_1,R_2 \to 0$ and
$R_3 \to \infty$, where the quartet structure of all types of branes guarantees
again all local tadpole cancellations without the need of any further 
splittings.


\section{Other models with one $D5$ brane}

In this Section we provide a description of other classes of
shift models with one $D5$ brane. For the sake of
brevity, we again display only direct-channel amplitudes.

\subsection{One winding shift and one momentum shift}

This case corresponds to introducing a winding shift $w_1$ along the
first torus, and a momentum shift $p_2$ along the second torus in
the table  $\sigma_2$ of eq. (\ref{sigmas}).
The resulting models contain $D9$ and $D5_1$ branes.
The direct-channel Klein bottle amplitude is
\be
{\cal K} \!\!=\!\! \frac{1}{8} \biggl\{ T_{oo}\biggl[P_1P_2P_3 \!+\! P_1W_2W_3 
\!+\! W_1(-1)^{m_2}P_2W_3 \!+\! (-1)^{n_1}W_1W_2P_3\biggr]
\!+\! 2 \times 16 T_{go}P^{m+1/2}_1\left(\frac{ \eta}{\theta_4}\right)^2 
\biggr\} \ ,
\label{u1}
\ee
the direct-channel annulus amplitude is
\ba
{\cal A} &=& \frac{1}{8} \biggl\{ 
T_{oo}\biggr[\frac{N^2}{2}(P_1+P_1^{m+1/2})P_2P_3 
+ \frac{D_1^2}{4}(P_1+P_1^{m+1/2})(W_2+W_2^{n+1/2})W_3 \biggr] \nonumber \\
&+& 2F^2T_{of}(-1)^{m_2}P_2 \left( \frac{2\eta}{\theta_2} \right)^2
+ T_{go} ND_1(P_1^{m+1/4} + P_1^{m+3/4})\left( \frac{\eta}{\theta_4}\right)^2
 \biggr\}
\quad ,
\label{u2}
\ea
and finally the M\"obius projection is
\ba
{\cal M} &=& - \frac{1}{8} \biggl\{ 
\hat{T}_{oo} \biggl [NP_1P_2P_3 + D_1 P_1W_2W_3\biggr]
- \hat{T}_{og}(D_1 + N)P_1^{m+1/2} \left( \frac{2 \hat{\eta}}{\hat{\theta_2}}
\right)^2  \nonumber \\
&-& N \hat{T}_{of}(-1)^{m_2}P_2
\left( \frac{2 \hat{\eta}}{\hat{\theta_2}}\right)^2    \biggr\} \quad .
\label{u3}
\ea
\vskip 5pt
\input epsf \centerline{ \epsfbox{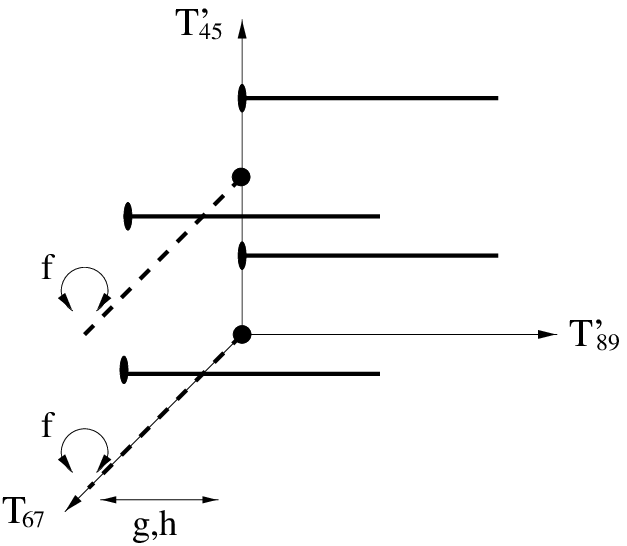}}
\begin{center}
Figure 6. $D_9$ (dashed) and $D5_1$ branes for
the $w_1p_2$ model.
\end{center}
\vskip 5pt

Figure 6 shows the brane configuration, after a T-duality along $T_{45}$
and $T_{89}$, that turns the $D9$ branes into $D5'_2$ branes (dashed), 
the $D5_1$ branes into $D5'_3$ branes.

We can parameterize the charges according to
\be
N = 2(n + \bar{n}) \ , \hspace{1cm} F = i(n - \bar{n}) \ , \hspace{1cm} 
D_1=4d \quad ,
\label{u4}
\ee
and the tadpole conditions 
\be
N = D_1 = 32  \ , 
\label{u5}
\ee
then require a gauge group
\be
U(8)_9 \times SO(8)_{5_1} \ .
\label{u6}
\ee
With a discrete Wilson line on the M\"obius strip, as in \cite{bs},
one can also
obtain a class of models with $D9$ gauge groups $SO(n) \times
SO(16-n)$.

The massless spectrum is again not chiral and has $N=2$ supersymmetry
on the $D9$ branes and $N=4$ supersymmetry on the
$D5_1$ branes. Actually, the whole massive spectrum of $D5_1$ branes
has $N=2$ supersymmetry.
Our thumb rule explains rather naturally these properties:
the $D9$ branes are not affected by the (parallel) winding
shift, while the $D5_1$ branes, orthogonal to both directions of
breaking (after $T_{45}$ duality) arrange themselves in a quartet
geometry, with $N=4$ supersymmetry. The multiplet structure of the
branes is, once more, directly compatible with all local tadpole
cancellations.
Aside from the gauge multiplets, the 
spectrum contains pairs of $N=2$ hypermultiplets in the
representation $(28,1)$ from the 
$99$ sector.

\subsection{One winding shift and two momentum shifts}

This case corresponds to introducing a winding shift $w_1$ along the
first torus, and two momentum shifts $p_2$ and $p_3$ along the other
two tori in
the table  $\sigma_2$ of eq. (\ref{sigmas}).
The resulting models contain $D9$ and $D5_1$ branes.
The direct-channel Klein bottle amplitude is
\ba
{\cal K} &=& \frac{1}{8} \biggl\{ T_{oo}\biggl[P_1P_2P_3 
+ P_1W_2W_3 + W_1(-1)^{m_2}P_2W_3 
+ (-1)^{n_1}W_1W_2(-1)^{m_3}P_3\biggr] \nonumber \\
&+& 2 \times 16 T_{go}P^{m+1/2}_1\left(\frac{ \eta}{\theta_4}\right)^2 
\biggr\} \quad ,
\label{u9}
\ea
the direct-channel annulus amplitude is
\ba
{\cal A} &=& \frac{1}{8} \biggl\{ 
T_{oo}\biggl[\frac{N^2}{2}(P_1+P_1^{m+1/2})P_2P_3 
+ \frac{D_1^2}{4}(P_1+P_1^{m+1/2})(W_2W_3+W_2^{n+1/2}W_3^{n+1/2})\biggr] 
\nonumber \\
&+& 2F^2 T_{of}(-1)^{m_2}P_2\left( \frac{2\eta}{\theta_2} \right)^2 
+ T_{go} ND_1 (P_1^{m+1/4}+P_1^{m+3/4})\left( \frac{\eta}{\theta_4} \right)^2 
\biggr \} \ .
\label{u10}
\ea
and finally the M\"obius projection is
\ba
{\cal M} &=& - \frac{1}{8} \biggl\{ \hat{T}_{oo}\biggl[NP_1P_2P_3 +
D_1P_1W_2W_3 \biggr]
- \hat{T}_{og}(D_1+N)P_1^{m+1/2}
\left( \frac{2 \hat{\eta}}{\hat{\theta_2}}\right)^2 \nonumber \\
&-& N \hat{T}_{of}(-1)^{m_2}
P_2 \left( \frac{2 \hat{\eta}}{\hat{\theta_2}}\right)^2   
\biggr\} \quad .
\label{u11}
\ea
\vskip 5pt
\input epsf \centerline{ \epsfbox{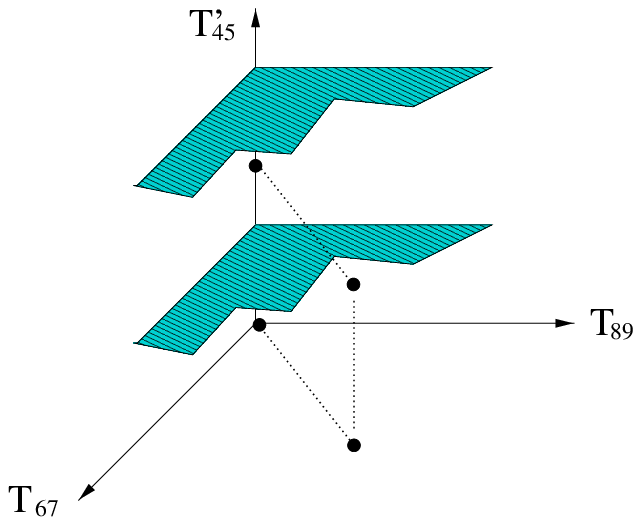}}
\begin{center}
Figure 7. $D_9$ and $D5_1$ branes for
the $w_1p_2p_{3}$ model, after $T_{45}$ duality.
\end{center}
\vskip 5pt
The figure shows, after a $T_{45}$ duality, the configuration of the 
$D9$ branes (now $D7$ branes, represented as two planes), and of the 
$D5_1$ branes (now $D3$ branes) in the model.

Also in this case, we can parameterize the charges according to
\be
N = 2(n + \bar{n}) \ , \hspace{1cm} F = i(n - \bar{n}) \ , \hspace{1cm} 
D_1=4d \quad ,
\label{u12}
\ee
and the tadpole conditions 
\be
N = D_1 = 32 \ ,  
\label{u13}
\ee
then require a gauge group
\be
U(8)_9 \times SO(8)_{5_1} \ .
\label{u14}
\ee
Again, with a discrete Wilson line on the M\"obius strip, one can also
obtain a class of models with $D9$ gauge groups $SO(n) \times
SO(16-n)$.

The massless spectrum is again not chiral and has $N=2$ supersymmetry
on the $D9$ branes and $N=4$ supersymmetry on the
$D5_1$ branes. Moreover, the whole massive spectrum of
$D5_1$ branes has $N=2$ supersymmetry.
Aside from the gauge multiplets, the 
spectrum contains pairs of $N=2$ hypermultiplets in the
representation $(28,1)$ from the 
$99$ sector.
In this model the $D9$ brane is orthogonal to the $R_1$
direction, while the $D5_1$ brane is orthogonal to all three
directions used for the breaking. Therefore, the complete local
tadpole cancellation in the simultaneous limit $R_1 \to 0$, $R_2,R_3
\to \infty$ requires a further splitting for the $D5_1$ brane, with
a resulting gauge group $U(8)_9 \times {[ SO(4) \times SO(4)]}_{5_1}$.

\subsection{One momentum shift and two winding shifts}

This case corresponds to introducing a momentum shift $p_2$ along the
second torus, and two winding shifts $w_1$ and $w_3$ along the other
two tori in the table  $\sigma_1$ of eq. (\ref{sigmas}).
The resulting models contain $D9$ and $D5_1$ branes.
The direct-channel Klein bottle amplitude is
\ba
{\cal K} &=& \frac{1}{8} \biggl\{ 
T_{oo}\biggl[P_1P_2P_3 + P_1W_2W_3 
+ W_1(-1)^{m_2}P_2(-1)^{n_3}W_3 + (-1)^{n_1}W_1W_2P_3\biggr] \nonumber \\
&+& 2 \times 16 T_{go}P^{m+1/2}_1\left(\frac{ \eta}{\theta_4}\right)^2
\biggr\} \quad ,
\label{u17}
\ea
the direct-channel annulus amplitude is
\ba
{\cal A} &=& \frac{1}{8} \biggl\{ 
T_{oo}\biggl[\frac{N^2}{4}(P_1+P_1^{m+1/2})P_2(P_3+P_3^{m+1/2}) 
+ \frac{D_1^2}{4}(P_1+P_1^{m+1/2})(W_2+W_2^{n+1/2})W_3\biggr] \nonumber \\
&+& T_{go} ND_1 (P_1^{m+1/4}+P_1^{m+3/4}) 
\left( \frac{\eta}{\theta_4} \right)^2
\biggr\} \quad ,
\label{u18}
\ea
and finally the M\"obius projection is
\be
{\cal M} = - \frac{1}{8} \biggl\{ \hat{T}_{oo}\biggl[N P_1P_2P_3 
+ D_1 P_1W_2W_3\biggr] - \hat{T}_{og}(D_1+N)P_1^{m+1/2} 
\left( \frac{2 \hat{\eta}}{\hat{\theta_2}}\right)^2  
\biggr\} \ .
\label{u19}
\ee
\vskip 5pt
\input epsf \centerline{ \epsfbox{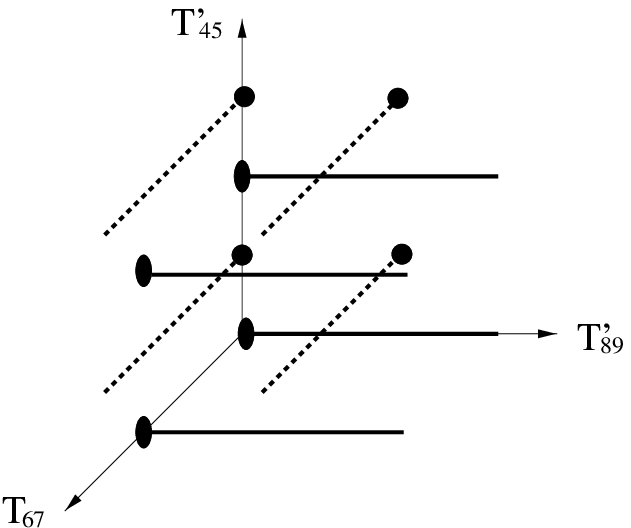}}
\begin{center}
Figure 8. $D_9$ (dashed) and $D5_1$ branes for
the $w_1p_2w_3$ model, after $T_{45}$ and $T_{89}$ dualities.
\end{center}
\vskip 5pt
Letting  $N=4n$ and $D_1=4d$, the tadpole conditions result in the
gauge group
\be
SO(8)_9 \times SO(8)_{5_1} \quad .
\label{u20}
\ee
In this model the interesting limits are $R_1,R_3 \to 0$ and $R_2 \to
\infty$. The D9 branes are orthogonal to the $R_1,R_3$ directions, while
the $D5_1$ branes are orthogonal to the $R_1,R_2$ directions. Thus, complete
local tadpole cancellations ask for quadruplets of D9 and D5 branes, 
that is indeed the configuration shown in
Figure 8. The resulting massless spectrum has $N=4$ supersymmetry,
both on the $D9$ and on the $D5_1$ branes, while the massive spectrum
has $N=2$ supersymmetry on both.

\section{Freely-acting orbifold models without $D5$ branes}

In this Section we provide a description of the remaining classes of
shift models, that do not contain $D5$ branes. As in the previous
Sections, for the sake of
brevity we only display direct-channel amplitudes.

\subsection{One momentum shift and
two winding shifts}

This case corresponds to introducing a momentum shift $p_1$ along the
first torus and two winding shifts $w_2$ and $w_3$ along the other tori in
the table $\sigma_2$ of eq. (\ref{sigmas}).
The direct-channel Klein bottle amplitude is
\be
{\cal K} = \frac{1}{8}  
T_{oo}\biggl[P_1P_2P_3 + (-1)^{m_1}P_1W_2W_3 + W_1P_2(-1)^{n_3}W_3 
+ W_1(-1)^{n_2}W_2P_3\biggr] \quad ,
\label{z1}
\ee
the direct-channel annulus amplitude is
\be
{\cal A} = \frac{1}{8} \biggl \{ T_{oo}\frac{N^2}{2}P_1(P_2P_3 
+ P_2^{m+1/2}P_3^{m+1/2})+2G^2T_{og}(-1)^{m_1}P_1
\left( \frac{2\eta}{\theta_2} \right)^2
 \biggr\}   \quad ,
\label{z2}
\ee
and finally the M\"obius projection is
\be
{\cal M} = - \frac{1}{8} \biggl \{
\hat{T}_{oo}NP_1P_2P_3-  \hat{T}_{og}N(-1)^{m_1}P_1
\left( \frac{2 \hat{\eta}}{\hat{\theta_2}}\right)^2   \biggr \} \quad .
\label{z3}
\ee 

Also in this case, we can parameterize the charges according to
\be
N = 2(n + \bar{n}) \ , \hspace{1cm} G = i(n - \bar{n}) \quad ,
\label{z4}
\ee
and the tadpole conditions 
\be
N = 32 \ , \hspace{1cm} G = 0 \ . 
\label{z5}
\ee
then require a gauge group
\be
U(8)_9 \ .
\label{z6}
\ee
Again, with a discrete Wilson line on the M\"obius strip, one can also
obtain a class of models with $D9$ gauge groups $SO(n) \times
SO(16-n)$.
The spectrum is again not chiral and has $N=2$ supersymmetry at all
mass levels.
Aside from the gauge multiplets, the spectrum contains pairs of 
$N=2$ hypermultiplets in the representation $(28,1)$ from the 
$99$ sector.

The interesting limits to consider in this case are $R_1 \to \infty$ and
$R_2,R_3 \to 0$, and the D9 branes are orthogonal to $R_2$ and $R_3$. 
In the simultaneous $R_2,R_3 \rightarrow 0$ limit, local tadpole conditions
ask for a quadruplet structure, while the D9 branes have actually a
doublet structure. Therefore, the simultaneous local tadpoles ask for a
further breaking $U(8) \to U(4) \times U(4)$, while of course the
separate limits $R_2 \to 0$ (or  $R_3 \to 0$) are not singular.

\subsection{Three winding shifts}

This case corresponds to introducing three winding shifts $w_1$,
$w_2$ and  $w_3$ along the three tori in the table $\sigma_1$ 
of eq. (\ref{sigmas}).
The direct-channel Klein bottle amplitude is
\be
{\cal K} = \frac{1}{8}  T_{oo}\biggl [P_1P_2P_3 + P_1(-1)^{n_2}W_2W_3 
+ W_1P_2 (-1)^{n_3}W_3 + (-1)^{n_1}W_1W_2P_3 \biggr ] \quad  ,
\label{z7}
\ee
the direct-channel annulus amplitude is
\be
{\cal A} = T_{oo}\frac{N^2}{32} \biggl [P_1P_2P_3 
+ P_1P_2^{m+1/2}P_3^{m+1/2} 
+ P_1^{m+1/2} P_2^{m+1/2}P_3+ P_1^{m+1/2}P_2P_3^{m+1/2} \biggr ]   \quad ,
\label{z8}
\ee
and finally the M\"obius projection is
\be
{\cal M} = - \frac{1}{8}\hat{T}_{oo}NP_1P_2P_3 \ .
\label{z9}
\ee 
Letting $N = 4n$ the tadpole conditions result in the gauge group
$SO(8)_9$ and the full spectrum has $N=4$ supersymmetry. 
The D9 branes are orthogonal to all three directions used for the
breaking and the relevant limits to study here are $R_1,R_2,R_3 \to
0$. All local tadpole conditions are automatically satisfied, thanks to the
quadruplet structure of branes, with the exception of the simultaneous
limit $R_1,R_2,R_3 \to 0$, where new tadpoles ask for the
further breaking $SO(8) \to SO(4) \times SO(4)$.


\section{Conclusions}

In this paper we have investigated the open descendants of $Z_2
\times Z_2$ orbifolds where the orbifold twists are
accompanied by various shifts on momentum or winding
modes. These shifts result in
a generalization \cite{kk} of the field-theoretical Scherk-Schwarz
mechanism \cite{ss} for (partial) 
supersymmetry breaking. Therefore, in all our models the
partial
breaking of supersymmetry may be regarded as spontaneous:
extended supersymmetry
can be recovered as some radii tend to infinity (zero) in
the case of momentum (winding) shifts. We have encountered new
instances of the phenomenon noticed in \cite{ads, adds}: 
the massless spectrum of branes orthogonal to the breaking
direction is unaffected at tree level, and
therefore has typically extended supersymmetry. This, in its turn,
makes these models not chiral. We have actually found that in 
several cases the extended supersymmetry is a property of the full
spectrum, not only of the massless modes, as in the asymmetric
orbifolds studied in \cite{bg}. We have displayed some
models with $N=2 \to N=1$,  $N=4 \to N=2$,  
$N=4 \to N=1$ partial breakings, in
addition to the other types, with $N=4 \to N=0$, $N=2 \to N=0$ and 
$N=4 \to N=2$ breakings,
discussed in \cite{ads,adds}. In particular, in Section 3 we have
collected a number of general results on a class of these shifted orbifolds, 
to which several others may be related via various T-dualities and 
modifications of the $\Omega$ projection.

 These models display an interesting new geometric feature: since the
shifts typically move the fixed points of the orbifolds, branes that
would be at fixed points in the unshifted case are actually moved away from
them (or, more precisely, from fixed tori, in
these $Z_2 \times
Z_2$ models). This gives rise to multiplets of, say, $m$ image branes,
that are typically
interchanged by some of the orbifold transformations and are left
invariant by others. Thus, in general only some of the orbifold
projections affect the partition function, and as a result the massless
modes, and often even the massive ones,
have generically extended supersymmetry. The rank
of the corresponding gauge groups is also affected, and is reduced to
$16/m$. This phenomenon of brane displacement, discussed for the 
supersymmetric $T^4/Z_2$
model in \cite{gp}, presents some analogies with the
corresponding one
induced by quantized $NS-NS$ antisymmetric tensor field backgrounds in
the compact space \cite{Bab}. 

We have also seen how, in general, the shifts make some of tadpole 
terms in the Klein bottle massive, eliminating the
corresponding D5 branes, and we have presented explicit
realizations of
models with two, one
and no D5 branes in the spectrum. All these models are not chiral,
a feature that can be traced to the lack of some of the breaking
terms for $ND$ or $D_iD_j$ strings or, equivalently, to the
multiplet structure of the branes.
 
We have provided an M-theory interpretation of some of the models, 
that can be related to Scherk-Schwarz deformations along the eleventh
dimension of M-theory compactified on an appropriate manifold \cite{aq,
ads,adds}. In this case, once a local tadpole cancellation condition is
fulfilled in the coordinate used for the breaking, two sets of
branes orthogonal to the breaking direction may be interpreted as a
pair of Horava-Witten walls.

These models provide examples of partial supersymmetry breaking where
gravity can be decoupled from the brane dynamics. Thus, it would be
interesting to study in detail the structure of the effective field
theory and to compare it with the known realizations of partial
supersymmetry breaking \cite{psb}. Moreover, the issue of radiative 
corrections is quite interesting, and was partly addressed, for the
gauge couplings, in \cite{abd}. 
For branes parallel to the breaking direction, the corrections have a 
logarithmic dependence on the corresponding radius, while
for branes orthogonal to the breaking
coordinate, in models with
local tadpole cancellation, they are exponentially small, in agreement
with the considerations presented in \cite{ab}.
\vskip 24pt
\begin{flushleft}
{\large \bf Acknowledgments}
\end{flushleft} 

We are grateful to C. Angelantonj, K. Ray, and in particular
to E. Kiritsis and Ya.S. Stanev for
useful discussions. The work of G.D. was supported in part by EEC TMR
contract  ERBFMRX-CT96-0090, while the work of
A.S. was supported in part by CNRS. G.D. and A.S. would like to thank
the Centre the Physique Th\'eorique of the Ecole Polytechnique
for the kind hospitality extended to them during the course of this research.
\vfill\eject


\end{document}